\documentclass[12pt,a4paper]{article}
\usepackage{a4wide,cite}
\usepackage{amsmath,amssymb}
\usepackage{latexsym,graphicx,epsfig}

\newcommand{\half}{{\textstyle\frac{1}{2}}}

\newcommand{\threehalf}{{\textstyle\frac{3}{2}}}

\def\lsim{\mathrel{\rlap{\raise 2.5pt \hbox{$<$}}\lower 2.5pt\hbox{$\sim$}}}
\def\gsim{\mathrel{\rlap{\raise 2.5pt \hbox{$>$}}\lower 2.5pt\hbox{$\sim$}}}

\renewcommand{\Re}{{\rm Re\thinspace}}
\renewcommand{\Im}{{\rm Im\thinspace}}

\allowdisplaybreaks 

\begin{document}
\renewcommand{\thefootnote}{\fnsymbol{footnote}}
\newpage\normalsize
    \pagestyle{plain}
    \setlength{\baselineskip}{4ex}\par
    \setcounter{footnote}{0}
    \renewcommand{\thefootnote}{\arabic{footnote}}
\newcommand{\preprint}[1]{%
\begin{flushright}
\setlength{\baselineskip}{3ex} #1
\end{flushright}}
\renewcommand{\title}[1]{%
\begin{center}
    \LARGE #1
\end{center}\par}
\renewcommand{\author}[1]{%
\vspace{2ex}
{\Large
\begin{center}
    \setlength{\baselineskip}{3ex} #1 \par
\end{center}}}
\renewcommand{\thanks}[1]{\footnote{#1}}
\begin{flushright}
    \today
\end{flushright}
\vskip 0.5cm

\begin{center}
{\bf \Large
Consistency of the Two Higgs Doublet Model and \\[4mm]
CP violation in top production at the LHC}
\end{center}
\vspace{1cm}
\begin{center}
Abdul Wahab El Kaffas$^{a}$,
Wafaa Khater$^{b}$, 
Odd Magne Ogreid$^{c}$  {\rm and}
Per Osland$^{a}$
\end{center}
\vspace{1cm}
\begin{center}
$^a$ Department of Physics and Technology,
University of Bergen, Postboks 7803, \\
N-5020 Bergen, Norway \\
$^{b}$ Department of Physics, Birzeit University, Palestine \\
$^{c}$ Bergen University College, Bergen, Norway
\end{center}
\vspace{1cm}
\begin{abstract}
It is important to provide guidance on whether CP violation may be measurable
in top-quark production at the Large Hadron Collider.  The present work
extends an earlier analysis of the non-supersymmetric Two-Higgs-Doublet Model
in this respect, by allowing a more general potential. Also, a more
comprehensive study of theoretical and experimental constraints on the model
is presented.  Vacuum stability, unitarity, direct searches and electroweak
precision measurements severely constrain the model. We explore, at low
$\tan\beta$, the allowed regions in the multidimensional parameter space that
give a viable physical model.  This exploration is focused on the parameter
space of the neutral sector rotation matrix, which is closely related to the
Yukawa couplings of interest.  In most of the remaining allowed regions, the
model violates CP.  We present a quantitative discussion of a particular
CP-violating observable.  This would be measurable in semileptonically
decaying top and antitop quarks produced at the LHC, provided the number of
available events is of the order of a million.
\end{abstract}
\section{Introduction}
\setcounter{equation}{0}
The Two-Higgs-Doublet Model (2HDM) is attractive as one of the simplest
extensions of the Standard Model that admits additional CP violation
\cite{Lee:iz,Weinberg:1976hu,Branco:1985aq}.  This is an interesting
possibility, given the unexplained baryon asymmetry of the Universe
\cite{Riotto:1999yt,Dine:2003ax}, and the possibility of exploring relevant,
new physics at the LHC \cite{Ellis:2004nh}.  In particular, the model can lead
to CP violation in $t\bar t$ production, a process which has received
considerable theoretical attention
\cite{Schmidt:1992et,Bernreuther:1993hq,Atwood:2000tu}, since it will become
possible to severely constrain or even measure it.

CP violation can be induced in $t\bar t$ production at the one-loop level, by
the exchange of neutral Higgs bosons which are not eigenstates under CP.  This
effect is only large enough to be of experimental interest if the neutral
Higgs bosons are reasonably light, and have strong couplings to the top
quarks.

Within the 2HDM (II), where the top quark gets its mass from coupling to the
Higgs field $\Phi_2$ \cite{HHG} (see sect.~\ref{sect:Yukawa}), the condition
of having sizable $Ht \bar t$ couplings forces us to consider small values of
$\tan\beta$.  A first exploration of this limit was presented in
\cite{Khater:2003wq}.  In that paper, the general conditions for measurability
of CP violation in $gg\to t\bar t$ at the LHC \cite{Bernreuther:1993hq} were
found to be satisfied in a certain region of the 2HDM parameter space. In
addition to having small $\tan\beta$, in order to have a measurable signal
with a realistic amount of data (of the order of a million $t\bar t$ events),
it was found necessary that the lightest neutral Higgs boson be light,
and that the spectrum not be approximately degenerate.  In fact, it was found
that in the most favourable observable considered, the effect would not reach
the per mil level unless there is one and only one Higgs boson below the
$t\bar t$ threshold, and that $\tan\beta$ is at most of order unity.  We here
extend the analysis of \cite{Khater:2003wq} to the more general case, allowing
the most general quartic couplings in the potential.

At small $\tan\beta$, also certain Yukawa couplings to charged Higgs bosons
are enhanced. Such couplings contribute to effects that are known
experimentally to very high precision.  In particular, at low $\tan\beta$ the
$B_d^0$--$\bar B_d^0$ oscillation data and the effective $Zb\bar b$ coupling,
measured via $R_b$ \cite{Grant:1994ak,Eidelman:2004wy} severely constrain the
model, whereas the $b\to s\gamma$ data \cite{Gambino:2001ew} constrain it at
low $M_{H^\pm}$.  Furthermore, the high-precision measurement of the $W$ and
$Z$ masses, as expressed via $\rho$ \cite{Eidelman:2004wy} constrains the
splitting of the Higgs mass spectrum.  Unless there are cancellations, the
charged Higgs boson can not be very much heavier than the lightest neutral
one, and the lightest neutral one can not be far away from the mass scale of
the $W$ and the $Z$ \cite{Bertolini:1985ia}. Also, the lightest one is
constrained by the direct searches at LEP
\cite{Boonekamp:2004ae,Abbiendi:2000ug}. We shall here study the interplay of
these constraints, and estimate the amount of CP violation that may be
measurable at the LHC in selected favorable regions of the remaining parameter
space.

An important characteristic of the 2HDM (as opposed to the MSSM
\cite{Wess:1974tw,Fayet:1976cr,Dimopoulos:1981zb,Nilles:1983ge,Haber:1984rc})
is the fact that, at the level of the mathematics, the masses of the neutral
and the charged Higgs bosons are rather independent (see
sect.~\ref{sect:2HDMReview}). However, the experimental precision on
$\Delta\rho$ (see sect.~\ref{sect:LEP-rho}) forces the charged Higgs mass to
be comparable in magnitude to the neutral Higgs masses.  Another important
difference is that whereas small values of $\tan\beta$ are practically
excluded in the MSSM \cite{Abbiendi:2004ww}, in the 2HDM, which has more free
parameters, they are not.

For a recent comprehensive discussion of the experimental constraints on the
2HDM (though mostly restricted to the CP-conserving limit), see
\cite{Diaz:2002tp} and \cite{Cheung:2003pw}.  The latter study, which
considers the CP-conserving limit, concludes that the model is practically
excluded, with the muon anomalous moment being very constraining.  However,
the interpretation of the data is now considered less firm, and furthermore,
that study focuses on large $\tan\beta$, and is thus less relevant for the
present work.

We present in sect.~\ref{sect:2HDMReview} an overview of the 2HDM, with focus
on the approach of ref.~\cite{Khater:2003wq}, and outline the present
extensions. In sect.~\ref{sect:model} we discuss the model in more detail, in
particular the implications of stability and unitarity, and review the
conditions for having CP violation. In sect.~\ref{sect:model-constraints} we
discuss various experimental constraints on the model, with particular
attention to small values of $\tan\beta$.  In
sect.~\ref{sect:parameter-overview} we present an overview of allowed
parameter regions, also restricted to small $\tan\beta$.  In
sect.~\ref{sect:ttbar} we discuss the implications of the model for a
particular CP-violating observable involving the energies of positrons and
electrons from the decays of $t$ and $\bar t$ produced in gluon--gluon
collisions at the LHC. Sect.~\ref{sect:summary} contains a summary and
conclusions.
\section{Review of the Two-Higgs-Doublet Model}
\label{sect:2HDMReview}
\setcounter{equation}{0}
The 2HDM may be seen as an unconstrained version of the Higgs sector of the
MSSM. While at tree level the latter can be parametrized in terms of only two
parameters, conventionally taken to be $\tan\beta$ and $M_A$, the 2HDM has
much more freedom. In particular, the neutral and charged Higgs masses are
rather independent. 

Traditionally, the 2HDM is defined in terms of the potential. The parameters
of the potential (quartic and quadratic couplings) determine the masses of the
neutral and the charged Higgs bosons. Alternatively, and this is the approach
followed here and in ref.~\cite{Khater:2003wq}, one can take masses and mixing
angles as input, and determine parameters of the potential as derived
quantities. This approach highlights the fact that the neutral and charged
sectors are rather independent, as well as masses being physically more
accessible than quartic couplings. However, some choices of input will lead to
physically acceptable potentials, others will not. This way, the two sectors
remain correlated.

In addition, the 2HDM neutral sector may or may not lead to CP violation,
depending on the choice of potential. We shall here consider the so-called
Model II, where $u$-type quarks acquire masses from a Yukawa coupling to one
Higgs doublet, $\Phi_2$, whereas the $d$-type quarks couple to the other,
$\Phi_1$. This structure is the same as in the MSSM.
\subsection{The approach of ref.~\cite{Khater:2003wq}}
\label{sect:khater}
The amount of CP violation that can be measured in $t\bar t$ production was
related to the Higgs mass spectrum and other model parameters in
\cite{Khater:2003wq}.  In that paper, the Higgs potential studied was
parametrized as \cite{Ginzburg:2001ss}
\begin{eqnarray}
\label{Eq:pot-KO}
V&=&\frac{\lambda_1}{2}(\Phi_1^\dagger\Phi_1)^2
+\frac{\lambda_2}{2}(\Phi_2^\dagger\Phi_2)^2
+\lambda_3(\Phi_1^\dagger\Phi_1) (\Phi_2^\dagger\Phi_2)
+\lambda_4(\Phi_1^\dagger\Phi_2) (\Phi_2^\dagger\Phi_1) \nonumber \\
&&+\frac{1}{2}\left[\lambda_5(\Phi_1^\dagger\Phi_2)^2+{\rm h.c.}\right] 
\nonumber \\
&&-\frac{1}{2}\left\{m_{11}^2(\Phi_1^\dagger\Phi_1)
+\left[m_{12}^2 (\Phi_1^\dagger\Phi_2)+{\rm h.c.}\right]
+m_{22}^2(\Phi_2^\dagger\Phi_2)\right\}.
\end{eqnarray}
Expanding the Higgs-doublet fields as
\begin{equation}
\Phi_i
=
\left(\begin{array}{c}
\varphi_i^+\\[1mm] 
\frac{1}{\sqrt{2}}(v_i+\eta_i+i\chi_i)
\end{array}\right)
\end{equation}
and choosing phases of $\Phi_i$ such that $v_1$ and $v_2$ are both real
\cite{Branco}, it is convenient to define
$\eta_3=-\sin\beta\chi_1+\cos\beta\chi_2$ orthogonal to the neutral Goldstone
boson $G^0=\cos\beta\chi_1+\sin\beta\chi_2$.  In the basis $(\eta_1,\ \eta_2,\
\eta_3)$, the resulting mass-squared matrix ${\cal M}^2$ of the neutral sector,
can then be diagonalized to physical states $(H_1,\ H_2,\ H_3)$ 
with masses $M_1\le M_2\le M_3$, via a rotation matrix $R$:
\begin{equation}
\begin{pmatrix}
H_1 \\ H_2 \\ H_3
\end{pmatrix}
=R
\begin{pmatrix}
\eta_1 \\ \eta_2 \\ \eta_3
\end{pmatrix},
\end{equation}
satisfying
\begin{equation}
\label{Eq:cal-M}
R{\cal M}^2R^{\rm T}={\cal M}^2_{\rm diag}={\rm diag}(M_1^2,M_2^2,M_3^2),
\end{equation}
and parametrized as\footnote{In ref.~\cite{Khater:2003wq}, these angles were
referred to as 
$(\tilde\alpha,\alpha_b,\alpha_c)\leftrightarrow(\alpha_1,\alpha_2,\alpha_3)$.}
\begin{align}     \label{Eq:R-angles}
R=R_3\,R_2\,R_1
=&\begin{pmatrix}
1         &    0         &    0 \\
0 &  \cos\alpha_3 & \sin\alpha_3 \\
0 & -\sin\alpha_3 & \cos\alpha_3
\end{pmatrix}
\begin{pmatrix}
\cos\alpha_2 & 0 & \sin\alpha_2 \\
0         &       1         & 0 \\
-\sin\alpha_2 & 0 & \cos\alpha_2
\end{pmatrix}
\begin{pmatrix}
\cos\alpha_1 & \sin\alpha_1 & 0 \\
-\sin\alpha_1 & \cos\alpha_1 & 0 \\
0         &       0         & 1
\end{pmatrix}  \nonumber \\
=&\begin{pmatrix}
c_1\,c_2 & s_1\,c_2 & s_2 \\
- (c_1\,s_2\,s_3 + s_1\,c_3) 
& c_1\,c_3 - s_1\,s_2\,s_3 & c_2\,s_3 \\
- c_1\,s_2\,c_3 + s_1\,s_3 
& - (c_1\,s_3 + s_1\,s_2\,c_3) & c_2\,c_3
\end{pmatrix}
\end{align}
with $c_i=\cos\alpha_i$, $s_i=\sin\alpha_i$.
The rotation angle $\alpha_1$ is chosen such that in the limit of 
no CP violation
($s_2\to0$, $s_3\to0$) then $\alpha_1\to\alpha+\half\pi$,
where $\alpha$ is the familiar mixing angle of the CP-even sector,
and the additional $\half\pi$ provides the mapping $H_1\leftrightarrow h$,
instead of $H$ being in the $(1,1)$ position of ${\cal M}^2_{\rm diag}$,
as is used in the MSSM \cite{HHG}.

While the signs of $\eta_i$ and $\chi_i$ are fixed by our choice of taking the
vacuum expectation values real and positive \cite{Branco}, the phase of $H_i$
has no physical consequence. One may therefore freely change the sign of one
or more rows, e.g., let $R_{1i}\to-R_{1i}$ (see
sect.~\ref{sect:rotation-symmetries}).

Rather than describing the phenomenology in terms of the parameters of the
potential (\ref{Eq:pot-KO}), in \cite{Khater:2003wq} the physical mass of the
charged Higgs boson, as well as those of the two lightest neutral ones, were
taken as input, together with the rotation matrix $R$.  Thus, the input can be
summarized as
\begin{equation}
\text{Parameters:\ }
\tan\beta,\ (M_1,\ M_2),\ (M_{H^\pm},\ \mu^2),\ (
\alpha_1,\ \alpha_2,\ \alpha_3),
\end{equation}
where $\tan\beta=v_2/v_1$ and $\mu^2=v^2\nu$, with $\nu=\Re
m_{12}^2/(2v_1v_2)$ and $v=246~\text{GeV}$.

This approach provides better control of the physical content of the model.
In particular, the elements $R_{13}$ and $R_{23}$ of the rotation matrix must
be non-zero in order to yield CP violation. For consistency, this requires
$\Im\lambda_5$ and $\Im m^2_{12}$ (as derived quantities) to be non-zero.

\subsection{The general potential}
For the potential, in this study, we take
\begin{equation}
\label{Eq:pot_7}
V=\text{Expression}~(\ref{Eq:pot-KO})
+\left\{\left[\lambda_6(\Phi_1^\dagger\Phi_1)+\lambda_7
(\Phi_2^\dagger\Phi_2)\right](\Phi_1^\dagger\Phi_2)
+{\rm h.c.}\right\}.
\end{equation}
The new terms proportional to $\lambda_6$ and $\lambda_7$ have to be
carefully constrained, since this potential does not satisfy
natural flavour conservation \cite{Glashow:1976nt}, even if each
doublet is coupled only to up-type or only to down-type flavours.

The various coupling constants in the potential will of course depend on the
choice of basis $(\Phi_1,\Phi_2)$. Recently, there has been some focus
\cite{Davidson:2005cw} on the importance of formulating physical observables
in a basis-independent manner. Here, we shall adopt the so-called Model~II
\cite{HHG} for the Yukawa couplings. This will uniquely identify the basis in
the $(\Phi_1,\Phi_2)$ space.

Minimizing the potential (\ref{Eq:pot_7}), we can rewrite it (modulo a
constant) as
\begin{eqnarray}
V&=&\frac{\lambda_1}{2}\left[(\Phi_1^\dagger\Phi_1)-\frac{v_1^2}{2}\right]^2
+\frac{\lambda_2}{2}\left[(\Phi_2^\dagger\Phi_2)
-\frac{v_2^2}{2}\right]^2 
+\lambda_3(\Phi_1^\dagger\Phi_1)(\Phi_2^\dagger\Phi_2)
+\lambda_4(\Phi_1^\dagger\Phi_2)(\Phi_2^\dagger\Phi_1)\nonumber \\
&&+\left\{\frac{1}{2}\lambda_5(\Phi_1^\dagger\Phi_2)^2
+\left[\lambda_6(\Phi_1^\dagger\Phi_1)
+\lambda_7(\Phi_2^\dagger\Phi_2)\right]
(\Phi_1^\dagger\Phi_2)
+{\rm h.c.}\right\} \nonumber \\
&&-\frac{1}{2}\bigl[\Re \lambda_{34567}-2\nu\bigr] 
[v_2^2(\Phi_1^\dagger\Phi_1) + v_1^2(\Phi_2^\dagger\Phi_2)] 
-v_1v_2\,\Re[\lambda_6(\Phi_1^\dagger\Phi_1)
+ \lambda_7(\Phi_2^\dagger\Phi_2)] \nonumber \\
&&- v_1v_2\bigl[2\nu\,\Re(\Phi_1^\dagger\Phi_2)
- \Im\lambda_{567}\,\Im(\Phi_1^\dagger\Phi_2)\bigr].
\end{eqnarray}
Here and in the following, we adopt the abbreviations
\begin{equation}
\lambda_{345}
=\lambda_3+\lambda_4+\lambda_5, \quad
\lambda_{34567}
=\lambda_{345}
+\frac{v_1}{v_2}\,\lambda_6+\frac{v_2}{v_1}\,\lambda_7, \quad
\lambda_{567}
=\lambda_5+\frac{v_1}{v_2}\,\lambda_6+\frac{v_2}{v_1}\,\lambda_7.
\end{equation}

The mass-squared matrix ${\cal M}^2$ of (\ref{Eq:cal-M}), corresponding to the
neutral sector of the potential, is found to be
\begin{eqnarray}
\label{Eq:M_ij}
{\cal M}^2_{11}&=&v^2[c_\beta^2\,\lambda_1 +s_\beta^2\,\nu 
+\frac{s_\beta}{2c_\beta}\Re(3c_\beta^2\,\lambda_6-s_\beta^2\,\lambda_7)], 
\nonumber \\
{\cal M}^2_{22}&=&v^2[s_\beta^2\,\lambda_2+c_\beta^2\,\nu
+\frac{c_\beta}{2s_\beta}\Re(-c_\beta^2\,\lambda_6+3s_\beta^2\,\lambda_7)], 
\nonumber \\
{\cal M}^2_{33}&=&v^2\Re[-\lambda_5+\nu-\frac{1}{2c_\beta s_\beta} 
(c_\beta^2\,\lambda_6 +s_\beta^2\,\lambda_7)], \nonumber \\
{\cal M}^2_{12}&=&v^2[c_\beta s_\beta(\Re\lambda_{345}-\nu)
+\threehalf\Re(c_\beta^2\,\lambda_6+s_\beta^2\,\lambda_7)], \nonumber \\
{\cal M}^2_{13}&=&-\half v^2\Im[s_\beta\,\lambda_5+2c_\beta\,\lambda_6], 
\nonumber \\
{\cal M}^2_{23}&=&-\half v^2\Im[c_\beta\,\lambda_5+2s_\beta\,\lambda_7],
\end{eqnarray}
with ${\cal M}^2_{ji}={\cal M}^2_{ij}$.

Here, compared with the potential (\ref{Eq:pot-KO}), we have two more complex
parameters, $\lambda_6$ and $\lambda_7$ (four new real parameters), but rather
than those, we take as additional parameters $M_3$, $\Im\lambda_5$,
$\Re\lambda_6$ and $\Re\lambda_7$.  Thus, the input will be
\begin{equation}
\label{Eq:free-parameters}
\text{Parameters:\ }
\tan\beta,\ (M_1,\ M_2,\ M_3),\ (M_{H^\pm},\ \mu^2),\ 
(\alpha_1,\ \alpha_2,\ \alpha_3),\
\Im\lambda_5,\ (\Re\lambda_6,\ \Re\lambda_7).
\end{equation}

\section{Model properties}
\label{sect:model}
\setcounter{equation}{0}
We want to explore regions of parameter space where there is significant CP
violation.  In order to do that, we need to map out regions in the
$\{\alpha_1,\alpha_2,\alpha_3\}$ space where the model is consistent (figures
are presented in sect.~\ref{sect:parameter-overview}).

From eq.~(\ref{Eq:cal-M}), it follows that
\begin{equation}
\label{Eq:calM-RMsqR}
{\cal M}^2_{ij}=\sum_k R_{ki} M_k^2 R_{kj}.
\end{equation}
Here, it is evident that the signs of the rows of $R$ play no role.

Comparing the expressions (\ref{Eq:calM-RMsqR}) with (\ref{Eq:M_ij}), invoking
also
\begin{equation}
\label{Eq:mch}
M_{H^\pm}^2=\mu^2-\half v^2(\lambda_4+\Re\,\lambda_{567}),
\end{equation}
we can solve for the $\lambda$'s.  
In particular, it follows from (\ref{Eq:M_ij}) that
\begin{align} \label{Eq:Im-lam67}
\Im\lambda_6
&=-\frac{1}{c_\beta}\left[\frac{1}{v^2}{\cal M}^2_{13}
+\frac{s_\beta}{2}\Im\lambda_5\right], \nonumber \\
\Im\lambda_7
&=-\frac{1}{s_\beta}\left[\frac{1}{v^2}{\cal M}^2_{23}
+\frac{c_\beta}{2}\Im\lambda_5\right].
\end{align}
\subsection{Symmetries}
\label{sect:symmetries}
By exploiting certain symmetries of the rotation matrix $R$, we can reduce the
ranges of parameters that have to be explored.
\subsubsection{Transformations of the rotation matrix}
\label{sect:rotation-symmetries}
The rotation matrix $R$ is invariant
under the following transformation;
\begin{alignat}{5}  \label{Eq:symm-A}
&\text{A}:&\quad
\alpha_1\to\pi+\alpha_1,\;\alpha_2\to\pi-\alpha_2,\;
\alpha_3\to\pi+\alpha_3;
\end{alignat}
which leaves its elements unchanged.

Another class of transformations are those where two rows of 
$R$ (i.e., physical Higgs fields) change sign, as discussed in 
sect.~\ref{sect:khater}.
The transformations are \cite{Khater:2003wq}:
\begin{alignat}{5}  \label{Eq:symm-B}
&\text{B1}:&\quad 
&\alpha_1\to\pi+\alpha_1,\; \alpha_2\to\pi-\alpha_2,\;
\alpha_3\text{ fixed}: &\quad
R_{1i}&\to R_{1i}, &\quad R_{2i}&\to -R_{2i}, &\quad R_{3i}&\to -R_{3i},
\nonumber\\
&\text{B2}:&\quad
&\alpha_1\text{ fixed},\;\alpha_2\to\pi+\alpha_2,\;
\alpha_3\to-\alpha_3: &\quad
R_{1i}&\to -R_{1i}, &\quad R_{2i}&\to R_{2i}, &\quad R_{3i}&\to -R_{3i},
\nonumber \\
&\text{B3}:&\quad 
&\alpha_1\to\pi+\alpha_1,\; \alpha_2\to-\alpha_2,\;
\alpha_3\to-\alpha_3: &\quad
R_{1i}&\to -R_{1i}, &\quad R_{2i}&\to -R_{2i}, &\quad R_{3i}&\to R_{3i}.
\end{alignat}
Actually, any one of these is a combination of the other two.
For example, the transformation B3 is the combination of B1 and B2.
Other transformations exist that will yield the same symmetries, but
they will be combinations of one of these three transformations 
followed by the transformation A. 
In total we have 6 different transformations that yield symmetries 
of type B.

The third class of transformation we consider are those where two columns of
$R$ change sign.
These transformations are:
\begin{alignat}{5}  \label{Eq:symm-C}
&\text{C1}:&\quad 
&\alpha_1\to\pi-\alpha_1,\;\alpha_2\to\pi+\alpha_2,\;
\alpha_3\text{ fixed}: &\quad
R_{j1}&\to R_{j1}, &\quad R_{j2}&\to -R_{j2}, &\quad R_{j3}&\to -R_{j3},
\nonumber\\
&\text{C2}:&\quad
&\alpha_1\to-\alpha_1,\; \alpha_2\to\pi+\alpha_2,\;
\alpha_3\text{ fixed}: &\quad
R_{j1}&\to -R_{j1}, &\quad R_{j2}&\to R_{j2}, &\quad R_{j3}&\to -R_{j3},
\nonumber\\
&\text{C3}:&\quad 
&\alpha_1\to\pi+\alpha_1,\;\alpha_2\text{ fixed},\;
\alpha_3\text{ fixed}: &\quad
R_{j1}&\to -R_{j1}, &\quad R_{j2}&\to -R_{j2}, &\quad R_{j3}&\to R_{j3}.
\end{alignat}
The transformation C3 is the combination of the transformations 
C1 and C2.
Other transformations exist that will yield the same symmetries, but
they will be combinations of one of these three transformations 
followed by the transformation A. 
In total we have 6 different transformation that yield symmetries 
of type C.

Under transformations of type A and B, the resulting mass-squared matrix 
${\cal M}^2=R^T{\cal M}_{\rm diag}^2R$ will be invariant. We make use of
this fact along with the symmetries A, B1 and B2 to reduce the parameter 
space under consideration to 
\begin{equation} \label{Eq:angular-range}
-\pi/2<\{\alpha_1,\alpha_2,\alpha_3\}\leq\pi/2.
\end{equation}

Under transformations of type C, the mass-squared matrix will not be invariant,
some of its non-diagonal elements will change sign while the rest 
are unaltered.
\begin{alignat}{5}  \label{Eq:symm-C-M}
&\text{C1}:&\quad 
{\cal M}^2_{12}&\to -{\cal M}^2_{12}, &\quad
{\cal M}^2_{13}&\to -{\cal M}^2_{13}, &\quad
{\cal M}^2_{23}&\to +{\cal M}^2_{23},
\nonumber\\
&\text{C2}:&\quad
{\cal M}^2_{12}&\to -{\cal M}^2_{12}, &\quad
{\cal M}^2_{13}&\to +{\cal M}^2_{13}, &\quad
{\cal M}^2_{23}&\to -{\cal M}^2_{23}, &\quad
\nonumber\\
&\text{C3}:&\quad 
{\cal M}^2_{12}&\to +{\cal M}^2_{12}, &\quad
{\cal M}^2_{13}&\to -{\cal M}^2_{13}, &\quad
{\cal M}^2_{23}&\to -{\cal M}^2_{23}.
\end{alignat}

While a change of the sign of ${\cal M}^2_{12}$ implies changes in the
physical content of the model, a change of sign of ${\cal M}^2_{13}$ and/or
${\cal M}^2_{23}$ can be compensated for by adjusting the imaginary parts of
$\lambda_5$, $\lambda_6$ and $\lambda_7$.  Thus, the most interesting
transformation among the set (\ref{Eq:symm-C-M}) is C3.

The transformation $\text{B3}\cdot\text{C3}$ is physically
equivalent to C3 since transformations of type B leave the mass-squared 
matrix invariant:
\begin{alignat}{5} 
\text{B3$\cdot$C3}:&\quad \label{Eq:symm-B3-C3}
&\alpha_1\text{ fixed},\alpha_2\to-\alpha_2,\;
\alpha_3\to-\alpha_3: &\quad
{\cal M}^2_{12}&\to +{\cal M}^2_{12}, &\quad
{\cal M}^2_{13}&\to -{\cal M}^2_{13}, &\quad
{\cal M}^2_{23}&\to -{\cal M}^2_{23}.
\end{alignat}

When $\Im\lambda_5=0$, it follows from (\ref{Eq:Im-lam67}) that a sign change
of ${\cal M}^2_{13}$ and ${\cal M}^2_{23}$ can be compensated for by sign
changes of $\Im\lambda_6$ and $\Im\lambda_7$.  These signs play no role in the
discussion of stability (see Appendix~A) and unitarity \cite{Ginzburg:2003fe}.
We shall therefore, when discussing the case $\Im\lambda_5=0$
(sects.~\ref{sect:mass-parameters-setA} and \ref{sect:mass-parameters-setB}),
make use of (\ref{Eq:symm-B3-C3}) to restrict the angular range from
(\ref{Eq:angular-range}) to the smaller
\begin{equation} \label{Eq:angular-range-restricted}
-\pi/2<\{\alpha_1,\alpha_2\}\leq\pi/2, \quad
0\leq\alpha_3\leq\pi/2.
\end{equation}
When $\Im\lambda_5\neq0$ we need to consider the angular range as given in
(\ref{Eq:angular-range}).

\subsubsection{Inversion of $\tan\beta$}
\label{sect:inversion-symmetry}

The Higgs sector is invariant under
\begin{alignat}{2} \label{Eq:tanbeta-symm}
\tan\beta&\leftrightarrow\cot\beta, &\qquad
\alpha_1&\leftrightarrow\half\pi-\alpha_1, \nonumber \\
\alpha_2&\leftrightarrow-\alpha_2, &\qquad
\alpha_3&\leftrightarrow\alpha_3,
\end{alignat}
accompanied by
\begin{alignat}{2} \label{Eq:tanbeta-symm-part2}
\lambda_1&\leftrightarrow\lambda_2, &\qquad
\lambda_3,\lambda_4&\leftrightarrow\lambda_3,\lambda_4, \nonumber \\
\lambda_5&\leftrightarrow\lambda_5^\star, &\qquad
\lambda_6&\leftrightarrow\lambda_7^\star.
\end{alignat}
This is just the symmetry between $\Phi_1$ and $\Phi_2$, and will be violated
by the introduction of Model II Yukawa couplings, which distinguish between
the two Higgs doublets, i.e., between $\tan\beta$ and $\cot\beta$.

\subsection{CP violation}
In general, with all three rotation-matrix angles non-zero, the model will
violate CP. However, in certain limits, this is not the case.  In order {\it
not} to have CP violation, the mass-squared matrix must be block diagonal,
i.e., one must require
\begin{equation}
\label{Eq:no-CPv}
{\cal M}^2_{13}={\cal M}^2_{23}=0.
\end{equation}
Thus, CP {\it conservation} requires
\begin{align} \label{Eq:CP-conservation}
{\cal M}^2_{13}&=M_1^2R_{11}R_{13}+M_2^2R_{21}R_{23}+M_3^2R_{31}R_{33}=0,
\nonumber \\
{\cal M}^2_{23}&=M_1^2R_{12}R_{13}+M_2^2R_{22}R_{23}+M_3^2R_{32}R_{33}=0.
\end{align}
One possible solution of (\ref{Eq:CP-conservation}) is that
\begin{equation}
M_1=M_2=M_3.
\end{equation}
The expressions (\ref{Eq:CP-conservation}) then vanish, by the orthogonality
of $R$. There are additional limits of no CP violation, as discussed below.

Expressed in terms of the angles of the rotation matrix, the above elements
describing mixing of the CP-even and CP-odd parts of ${\cal M}^2$ take the
form
\begin{align}
{\cal M}^2_{13}&= c_1 c_2 s_2(M_1^2-s_3^2M_2^2-c_3^2M_3^2)
+ s_1 c_2 c_3 s_3(-M_2^2+M_3^2), \nonumber \\
{\cal M}^2_{23}&= c_1 c_2 c_3 s_3(M_2^2-M_3^2)
+ s_1 c_2 s_2(M_1^2-s_3^2M_2^2-c_3^2M_3^2).
\end{align}
In the mass-non-degenerate case, they vanish (there is thus no CP violation)
if either:
\begin{alignat}{2} \label{Eq:CPV-condition}
&\text{Case I:} &\quad
&\sin2\alpha_2=0, \quad \text{and} \quad \sin2\alpha_3=0, 
\quad\text{or} \nonumber \\
&\text{Case II:} &\quad
&\cos\alpha_2=0, \quad \alpha_3 \quad \text{arbitrary}.
\end{alignat}
Note that $M_1^2-s_3^2M_2^2-c_3^2M_3^2<0$ 
for non-degenerate or partially degenerate masses, ordered such that
$M_1\le M_2\le M_3$ (where no more than two of the masses are equal).
Thus, there are no additional CP-conserving solutions for the vanishing
of this factor. The cases of partial degeneracy, $M_1=M_2\ne M_3$,
and $M_1\ne M_2=M_3$ will be discussed in sect.~\ref{subsect:degeneracy}.

It is thus natural to focus on the angles $\alpha_2$ and $\alpha_3$.
In particular, since $R_{12}R_{13}$ is associated
with CP-violation in the $H_1t\bar t$ coupling (see sect.~\ref{sect:Yukawa}), 
we are interested in regions where $|\sin(2\alpha_2)|$ is large.

\subsection{Reference parameters}
In order to search for parameters with ``large'' CP violation, we will
assume $H_1$ is light, and that $M_2$ is not close to $M_1$, as such
degeneracy would cancel any CP violation.

For illustration, as a conservative default set of parameters, we take
\begin{equation}
\text{Set A:}\quad
\begin{cases}
\tan\beta=\{0.5,1.0,2.0\},\ (M_1,M_2,M_3)=(100,300,500)~\text{GeV}, \\
M_{H^\pm}=500~\text{GeV},\ \mu^2=(200~\text{GeV})^2,\
\Im\lambda_5=0,\ \Re\lambda_6=\Re\lambda_7=0.
\end{cases}
\end{equation}
Here, the lightest neutral Higgs boson can be accommodated by the negative LEP
searches \cite{Boonekamp:2004ae,Abbiendi:2000ug} provided it does not couple
too strongly to the $Z$, and the charged Higgs boson mass is compatible with
the negative LEP \cite{Eidelman:2004wy} and Fermilab searches
\cite{Abulencia:2005jd} as well as with the $B_d^0$--$\bar B_d^0$ oscillation,
$R_b$ constraints \cite{Cheung:2003pw} (see
sect.~\ref{sect-mh_ch-constraints}) and the $b\to s\gamma$ analysis at low
$\tan\beta$ \cite{Gambino:2001ew}.

As a second set of parameters, we take 
\begin{equation}
\text{Set B:}\quad
\begin{cases}
\tan\beta=\{0.5,1.0, 2.0\},\ (M_1,M_2)=(80,300)~\text{GeV},\ 
M_3=\{400,600\}~\text{GeV}\\
M_{H^\pm}=300~\text{GeV},\ \mu^2=\{0,(200~\text{GeV})^2\},\
\Im\lambda_5=0,\ \Re\lambda_6=\Re\lambda_7=0.
\end{cases}
\end{equation}
This set, which represents a light Higgs sector, is marginally in conflict
with data (the combination of charged-Higgs mass and $\tan\beta$ values
violate the $R_b$ constraints by up to $5\sigma$, see Table~\ref{table:R_b} in
sect.~\ref{sect:Rb}), but is chosen for a more ``optimistic'' comparison,
since it could give more CP violation due to a lower value of $M_1$ (which
enhances the loop integrals).

\subsection{Stability and unitarity}
A necessary condition we must impose on the model, is that the potential is
positive when $|\Phi_1|$ and $|\Phi_2|\to\infty$.  
This constraint, which is rather involved, is discussed in Appendix~A.
Two obvious conditions are that
\begin{equation}
\label{Eq:pos-simple}
\lambda_1>0, \qquad \lambda_2>0.
\end{equation}
In general, the additional stability constraint is that
$\lambda_3$ and $\lambda_4$ cannot be ``too large and negative'',
and that $|\lambda_5|$, $|\lambda_6|$ , $|\lambda_7|$
cannot be ``too large''.

Furthermore, we shall impose tree-level unitarity on the
Higgs-Higgs-scattering sector, as formulated in
\cite{Akeroyd:2000wc,Ginzburg:2003fe} (see also
ref.~\cite{Kanemura:1993hm}). This latter constraint is related to the
perturbativity constraint ($\lambda$'s not allowed ``too large'') adopted in
ref.~\cite{Khater:2003wq}, but actually turns out to be numerically more
severe.

\subsection{Yukawa couplings}
\label{sect:Yukawa}

With the above notation, and adopting the so-called Model~II \cite{HHG} for the
Yukawa couplings, where the down-type and up-type quarks are coupled only
to $\Phi_1$ and $\Phi_2$, respectively, the couplings can be
expressed (relative to the SM coupling) as
\begin{alignat}{2}  \label{Eq:H_itt}
&H_j  b\bar b: &\qquad 
&\frac{1}{\cos\beta}\, [R_{j1}-i\gamma_5\sin\beta R_{j3}], \nonumber \\
&H_j  t\bar t: &\qquad 
&\frac{1}{\sin\beta}\, [R_{j2}-i\gamma_5\cos\beta R_{j3}]
\equiv a+i\tilde a\gamma_5.
\end{alignat}
Likewise, we have for the charged Higgs bosons \cite{HHG}
\begin{alignat}{2}  \label{Eq:Yukawa-charged}
&H^+ b \bar t: &\qquad 
&\frac{ig}{2\sqrt2 m_W}\, 
[m_b(1+\gamma_5)\tan\beta+m_t(1-\gamma_5)\cot\beta], \nonumber \\
&H^-  t\bar b: &\qquad 
&\frac{ig}{2\sqrt2 m_W}\, 
[m_b(1-\gamma_5)\tan\beta+m_t(1+\gamma_5)\cot\beta].
\end{alignat}
With this Yukawa structure, the model is denoted as
the 2HDM (II).

The product of the $H_j t\bar t$ scalar and pseudoscalar
couplings,
\begin{equation}   \label{Eq:gamma_CP-i}
\gamma_{CP}^{(j)}=-a\,\tilde a
=\frac{\cos\beta}{\sin^2\beta}\, R_{j2}R_{j3}
\end{equation}
plays an important role in determining the amount of CP violation
in the top-quark sector.

As was seen in ref.~\cite{Khater:2003wq}, unless the Higgs boson is resonant
with the $t\bar t$ system, CP violation is largest for small Higgs masses. For
a first orientation, we shall therefore focus on the contributions of the
lightest Higgs boson, $H_1$.  (There will also be significant contributions
from the two heavier Higgs bosons, as discussed in sect.~\ref{sect:ttbar}.)
For the lightest Higgs boson, the coupling (\ref{Eq:H_itt}) becomes
\begin{equation}    \label{Eq:gamma_CP}
H_1 t \bar t: \qquad 
\frac{1}{\sin\beta}\,[\sin\alpha_1\cos\alpha_2
                   -i\gamma_5\cos\beta\sin\alpha_2], \quad
\text{with} \quad 
\gamma_{CP}^{(1)}=\half\,
\frac{\sin\alpha_1\sin(2\alpha_2)}{\tan\beta\sin\beta},
\end{equation}
where $\alpha_1$ and $\alpha_2$ are 
mixing angles of the Higgs mass matrix as defined by Eqs.~(\ref{Eq:cal-M})
and (\ref{Eq:R-angles}).

From (\ref{Eq:gamma_CP}), we see that low $\tan\beta$ are required for having
large CP violation in the top-quark sector. However, according to
(\ref{Eq:Yukawa-charged}), for low $\tan\beta$ the charged-Higgs Yukawa
coupling is also enhanced. Thus, for low $\tan\beta$, the $R_b$, $\Delta
M_{B_d}$ \cite{Grant:1994ak,Eidelman:2004wy} and $b\to s\gamma$ constraints
\cite{Gambino:2001ew} force $M_{H^\pm}$ to be high. For a quantitative
discussion, see sect.~\ref{sect-mh_ch-constraints}.

\subsection{CP violation in the Yukawa sector}

We shall in sect.~\ref{sect:ttbar} study CP violation in the process
\begin{equation}
pp\to t\bar tX \to e^+e^-X,
\end{equation}
focusing on the sub-process
\begin{equation}
gg\to t\bar t \to e^+e^-X.
\end{equation}
Let the CP violating quantity of that process be given by
\cite{Bernreuther:1993hq,Khater:2003wq}
\begin{equation} \label{Eq:CPV-ttbar}
\sum_{j=1}^3 R_{j2}\,R_{j3}\,f(M_j),
\end{equation}
where $f(M_j)$ is some function of the neutral Higgs mass $M_j$, in general
determined by loop integrals.

When the three neutral Higgs bosons are light, they will {\it all} contribute
to the CP-violating effects. In fact, in the limit of three mass-degenerate
Higgs bosons, the model may still be consistent in the sense that solutions
can be found in some regions of parameter space,
but the CP violation will cancel, since [cf.\ eq.~(\ref{Eq:gamma_CP-i})]
\begin{equation}  \label{Eq:orthogonal}
\sum_{j=1}^3\gamma_{CP}^{(j)}
=\frac{\cos\beta}{\sin^2\beta}\sum_{j=1}^3 R_{j2}\,R_{j3}=0
\end{equation}
due to the orthogonality of $R$. 

\subsection{Degenerate limits} \label{subsect:degeneracy}

The set of free parameters (\ref{Eq:free-parameters}) permits all three
neutral Higgs masses to be degenerate.  As discussed above, in this limit
there is no CP violation, by orthogonality of the rotation matrix $R$.
However, in contrast to the case of $\lambda_6=\lambda_7=0$ studied in
\cite{Khater:2003wq}, the partial degeneracies are non-trivial and may lead to
CP violation for certain choices of the angles $\alpha_i$:

\paragraph{\boldmath$M_1=M_2\equiv m\ne M_3$.}
In this limit, the elements of ${\cal M}^2$ that induce CP violation, are
\begin{align} \label{Eq:limit:M1=M2}
{\cal M}^2_{13}&=R_{31}R_{33}(M_3^2-m^2)
=c_2 c_3(s_1s_3-c_1s_2 c_3)
(M_3^2-m^2), \nonumber \\
{\cal M}^2_{23}&=R_{32}R_{33}(M_3^2-m^2)
=-c_2 c_3(c_1s_3+s_1s_2 c_3)
(M_3^2-m^2).
\end{align}
These both vanish, when the conditions (\ref{Eq:CPV-condition}) are satisfied,
or else, when
\begin{equation}
\cos\alpha_3=0, \quad \alpha_1, \alpha_2\quad \text{arbitrary}.
\end{equation}

By orthogonality, when the two lighter Higgs bosons are degenerate, the CP
violation (\ref{Eq:CPV-ttbar}) in the top-quark sector is proportional to
\begin{equation}
R_{32}\,R_{33}[f(M_3)-f(m)]\sim{\cal M}^2_{23}.
\end{equation}
Thus, even though the model violates CP in the limit $M_1=M_2\ne M_3$,
by for example having ${\cal M}^2_{13}\ne0$,
the top-quark sector would not violate CP at the one-loop level
unless ${\cal M}^2_{23}\sim R_{32}R_{33}\ne0$.

\paragraph{\boldmath$M_1\ne M_2=M_3\equiv M$.}
In this limit, the elements of ${\cal M}^2$ that induce CP violation are
\begin{align}
{\cal M}^2_{13}&=-R_{11}R_{13}(M^2-M_1^2)
=- c_1 c_2 s_2(M^2-M_1^2), \nonumber \\
{\cal M}^2_{23}&=-R_{12}R_{13}(M^2-M_1^2)
=-s_1 c_2 s_2(M^2-M_1^2).
\end{align}
We note that these both vanish for $\sin(2\alpha_2)=0$, meaning $\alpha_2=0$
or $\alpha_2=\pi/2$.  Thus, in the limit $M_1\ne M_2=M_3$ and $\alpha_2=0$,
but $\alpha_3$ {\it arbitrary}, the model does {\it not} violate CP, in
agreement with the results of \cite{Khater:2003wq}.

In this limit of the two heavier Higgs bosons being degenerate,
the CP violation in the top-quark sector is proportional to\footnote{Whereas
both these degenerate limits yield CP-violation in the $t$-quark
sector proportional to ${\cal M}^2_{23}$, the corresponding quantities
in the $b$-quark sector are proportional to ${\cal M}^2_{13}$.}
\begin{equation}
R_{12}\,R_{13}[f(M_1)-f(M)]\sim{\cal M}^2_{23}.
\end{equation}

In our parametrization, this is non-zero for
\begin{equation} \label{Eq:altilde-alb}
\sin\alpha_1\ne0, \quad \sin2\alpha_2\ne0,
\end{equation}
but with $\alpha_3$ arbitrary.

In the more constrained model discussed in \cite{Khater:2003wq}, the latter
limits of only two masses being degenerate do not exist. 
In that case, with $\lambda_6=\lambda_7=0$, a degeneracy
of two masses forces the third one to have that same value.

\section{Experimental model constraints at low $\tan\beta$}
\label{sect:model-constraints}
\setcounter{equation}{0}

It is convenient to split the experimental constraints on the 2HDM into two
categories. There are those involving only the charged Higgs boson, $H^{\pm}$,
and those also involving the neutral ones. The former, like the non-discovery
of a charged Higgs boson, the $b\to s\gamma$ constraint \cite{Gambino:2001ew},
and the $B_d^0$--$\bar B_d^0$ oscillations \cite{Cheung:2003pw} do not depend
on the rotation matrix $R$ and the amount of CP violation.  They are given by
$M_{H^\pm}$ and its coupling to quarks, (\ref{Eq:Yukawa-charged}), i.e.,
on $\tan\beta$. On the other hand, constraints involving the neutral ones
depend on the details of the couplings, i.e., they depend sensitively on the
rotation matrix $R$ as well as on the neutral Higgs mass spectrum.  We shall
first review the constraints that depend only on the charged Higgs sector.

In subsections \ref{sect:LEP-noHiggs}--\ref{sect:Rb} we discuss constraints on
the model that depend on the neutral sector. For the purpose of determining
these constraints, one has to generalize some predictions for the
CP-conserving case to the CP-violating case.  Eqs.~(\ref{Eq:C^2}),
(\ref{Eq:deltarho-HH}), (\ref{Eq:deltarho-HG}), (\ref{Eq:Delta-amu}) and
(\ref{Eq:Rb-Denner-mod}) are the results of such generalizations. In the
CP-conserving limit, $R_{13}=R_{23}=R_{31}=R_{32}=0$, and these expressions
simplify accordingly.

\subsection{Constraints on the charged-Higgs sector} 
\label{sect-mh_ch-constraints}

There are three important indirect constraints on the charged-Higgs sector:
the $B_d^0$--$\bar B_d^0$ oscillations, $R_b$ and $b\to s\gamma$.

The mass splitting in the neutral $B_d$ mesons is sensitive to contributions
from box diagrams with top quark and charged Higgs exchange
\cite{Abbott:1979dt,Athanasiu:1985ie,Glashow:1987qe,Geng:1988bq}, involving
the Yukawa couplings (\ref{Eq:Yukawa-charged}). Indeed, the diagrams with one
or two $H^\pm$ exchanges give contributions proportional to $(m_t\cot\beta)^2$
or $(m_t\cot\beta)^4$ multiplied by functions of $M_{H^\pm}$ that
for large $M_{H^\pm}$ behave like $1/M_{H^\pm}^{2}$. These
contributions to $\Delta M_{B_d}$ will constrain low values of $M_{H^\pm}$, in
particular at low values of $\tan\beta$.

While $\Delta M_{B_d}$ is known experimentally to considerable precision,
$\Delta M_{B_d}=(3.304\pm 0.046)\times 10^{-10}~\text{MeV}$
\cite{Eidelman:2004wy}, its theoretical understanding is more limited. The
largest theoretical uncertainty is related to the parameter combination
$f_B^2B_B$ of the hadronic matrix element.  This is only known to a precision
of 10--15\%.  Thus, we cannot exclude models which give predictions for
$\Delta M_{B_d}$ that deviate from the SM value by this order of magnitude,
even if this deviation is large compared to the experimental precision.

\begin{table}[ht]
\refstepcounter{table}
\label{table:Delta-M_b^0}
\addtocounter{table}{-1}
\begin{center}
\renewcommand{\tabcolsep}{.75em}
\begin{tabular}{|c|c|c|c|c|}
\hline 
$\tan\beta$& 0.5 & 1.0 & 2.0 \\
\hline
Set B: $M_{H^\pm}=300~\text{GeV}$ & 270\% & 45\% & 10\% \\
Set A: $M_{H^\pm}=500~\text{GeV}$ & 140\% & 24\% & 5\% \\
\hline
\end{tabular} 
\caption{2HDM contribution to $\Delta M_b^0$ for parameter sets A and B,
relative to its absolute value.}
\end{center}
\end{table}

In table~\ref{table:Delta-M_b^0} we show the contribution to $\Delta M_b^0$
that are due to the additional 2HDM fields, for the two parameter sets
considered.  It is clear that $\tan\beta=0.5$ is incompatible with the
experimental and theoretical constraints on $\Delta M_b^0$, whereas the
$\tan\beta=1.0$ case is marginal.

As mentioned above, the $b\to s\gamma$ constraints \cite{Gambino:2001ew} also
force $M_{H^\pm}$ to be high, in particular for low $\tan\beta$.  A recent
analysis arrived at the bound $M_{H^\pm}\ge 300~\text{GeV}$
\cite{Misiak:2006ab}.  However, at the very low values of $\tan\beta$
considered here, they are less severe than the $\Delta M_b^0$ and $R_b$
constraints. The experimental constraints on $R_b$ (see sect.~\ref{sect:Rb})
depend on the charged Higgs mass as well as on the neutral Higgs spectrum.
However, this constraint is for low values of $\tan\beta$ practically
independent of the neutral spectrum.

\subsection{Higgs non-discovery at LEP} 
\label{sect:LEP-noHiggs}

One might think that both parameter Set~A and Set~B would be in conflict with
the negative direct searches at LEP, because of the low values of
$M_1$. However, these bounds are marginally evaded by two facts which both
dilute the experimental sensitivity.  First, the $H_1ZZ$ coupling is
suppressed by the square of the Higgs-vector-vector coupling, which relative
to the Standard-Model coupling is
\begin{equation} \label{Eq:ZZH}
H_j ZZ: \qquad 
[\cos\beta R_{j1}+\sin\beta R_{j2}], \quad\text{for }j=1.
\end{equation}
For large values of $|\sin\alpha_1|$ (which is of interest in order to
maximise $\gamma_{CP}^{(1)}$ of (\ref{Eq:gamma_CP})), $R_{11}$ will be rather
small, and the second term in (\ref{Eq:ZZH}), proportional to $R_{12}$, takes
over. But this is suppressed by the factor $\sin\beta$. For some quantitative
studies of this suppression, which can easily be by a factor of 2 or more, see
Fig.~8 in \cite{Khater:2003wq}.  Secondly, the typical decay channel, $H_1\to
b\bar b$, is suppressed by the square of the Yukawa coupling,
Eq.~(\ref{Eq:H_itt}).  For small values of $\tan\beta$, this is approximately
$\cos^2\alpha_1\cos^2\alpha_2+\sin^2\beta\sin^2\alpha_2$. In the limits of
interest, both terms are small.

In the analysis of LEP data by DELPHI\footnote{There are such studies also by
the other LEP collaborations (see, for example \cite{Abbiendi:2000ug}), but
limited to the CP-conserving case.}, a channel-specific dilution factor $C^2$
is defined by \cite{Boonekamp:2004ae}
\begin{equation}
\sigma_{Z(h\to X)}=\sigma_{Zh}^\text{ew}\times C^2_{Z(h\to X)},
\end{equation}
where $\sigma_{Zh}^\text{ew}$ is the Standard Model cross section for a
particular Higgs mass $M$. In the 2HDM, for an $H_1$ decaying to $b \bar b$ or
$\tau \bar \tau$, the dilution is caused by the two effects discussed
above: There is a reduced coupling to the $Z$ boson [see (\ref{Eq:ZZH})] and a
modified (typically reduced) coupling to the $b\bar b$ (or $\tau \bar \tau$)
[see (\ref{Eq:H_itt})].  Thus, we take
\begin{equation} \label{Eq:C^2}
C^2_{Z(H_1\to b\bar b)}=[\cos\beta R_{11}+\sin\beta R_{12}]^2\,
\frac{1}{\cos^2\beta}\, [R_{11}^2+\sin^2\beta R_{13}^2],
\end{equation}
and consider as excluded parameter sets those where this quantity exceeds the
LEP bounds, roughly approximated as \cite{Boonekamp:2004ae}
\begin{equation} \label{Eq:C^2-values}
C^2_{Z(H_1\to b\bar b)}=0.2\quad\text{at 100~GeV},\quad
\text{and 0.1 at 80~GeV}.
\end{equation}
The last term in (\ref{Eq:H_itt}), involving $R_{13}$, is absent in the
CP-conserving case. However, at small $\tan\beta$, it has little
effect. Actually, similar results are obtained for both the $b\bar b$ and
$\tau\bar\tau$ channels. Presumably, when these are combined, a more strict
limit would be obtained.

It is instructive to consider this expression (\ref{Eq:C^2}) and the
corresponding constraints in three simple limits:
\begin{equation} \label{Eq:LEP-smalltanbeta}
\tan\beta\ll1:\qquad
C^2_{Z(H_1\to b\bar b)}\simeq\cos^4\alpha_1\cos^4\alpha_2\ll1.
\end{equation}
This requires either $\alpha_1\to\pm\pi/2$ or $\alpha_2\to\pm\pi/2$.
\begin{equation}
\tan\beta=1:\qquad
C^2_{Z(H_1\to b\bar b)}
\simeq[(\cos\alpha_1+\sin\alpha_1)\cos\alpha_2]^2
[\cos^2\alpha_1\cos^2\alpha_2+\half\sin^2\alpha_2]\ll1.
\end{equation}
This requires either $\alpha_2\to\pm\pi/2$ or $\alpha_1\to-\pi/4$ or
\{$\alpha_1\to\pm\pi/2$ and $\alpha_2\to0$\}.
\begin{equation} \label{Eq:LEP-largetanbeta}
\tan\beta\gg1:\qquad
C^2_{Z(H_1\to b\bar b)}\simeq\tan^2\beta
\sin^2\alpha_1\cos^2\alpha_2
[\cos^2\alpha_1\cos^2\alpha_2+\sin^2\beta\sin^2\alpha_2]\ll1.
\end{equation}
This requires $\alpha_1\to0$ or $\alpha_2\to\pm\pi/2$ or
\{$\alpha_1\to\pm\pi/2$ and $\alpha_2\to0$\}.

Furthermore, we note that at negative $\alpha_1$, the LEP bound is to some
extent evaded for small and medium $|\alpha_2|$ by cancellation among the two
terms in the $H_1ZZ$ coupling.

\subsection{The $\rho$-parameter constraint} 
\label{sect:LEP-rho}

A very important constraint coming from electroweak precision data, is the
precise determination of the $\rho$-parameter
\cite{Ross:1975fq,Einhorn:1981cy}.  The quantity
\begin{equation}
\Delta\rho\equiv\frac{1}{M_W^2}
\bigl[A_{WW}(q^2=0)-\cos^2\theta_W A_{ZZ}(q^2=0)\bigr]
\end{equation}
measures how much the $W$ and $Z$ self-energies can deviate from the
Standard-Model value, being zero at the tree level.
The experiments (mostly at LEP) have put
severe constraints on $\rho$ \cite{:2005em}:
\begin{equation}
\rho=1.0050\pm0.0010.
\end{equation}
The measured deviation from unity is accommodated within the Standard Model,
and mostly due to the heavy top quark.

In the 2HDM additional contributions arise \cite{Bertolini:1985ia}, which are
determined by the couplings to the $W$ and the $Z$ of the Higgs particles, and
by the mass splittings within the Higgs sector, as well as the mass splittings
with respect to the $W$ and $Z$ bosons. The simplified forms provided in
\cite{HHG} can easily be re-expressed in terms of the mass eigenvalues and the
elements $R_{jk}$ of the rotation matrix for the CP-violating basis. For the
Higgs--Higgs contribution, we find (the relevant couplings are given
in Appendix~B):
\begin{align} \label{Eq:deltarho-HH}
&A_{WW}^{HH}(0)-\cos^2\theta_W\, A_{ZZ}^{HH}(0)\nonumber\\
&=\frac{g^2}{64\pi^2}\sum_j\Bigl[
\{[\sin\beta R_{j1}-\cos\beta R_{j2}]^2 +R^2_{j3}\} 
F_{\Delta\rho}(M_{H^\pm}^2,M_j^2)
\nonumber \\
&\quad -\sum_{k>j}[(\sin\beta R_{j1}-\cos\beta R_{j2})R_{k3}
-(\sin\beta R_{k1}-\cos\beta R_{k2})R_{j3}]^2\,
F_{\Delta\rho}(M_j^2,M_k^2) \Bigr],
\end{align}
where
\begin{equation}
F_{\Delta\rho}(m_1^2,m_2^2)=\half(m_1^2+m_2^2)
-\frac{m_1^2m_2^2}{m_1^2-m_2^2}\log\frac{m_1^2}{m_2^2}.
\end{equation}

For the Higgs--ghost contribution, we have to subtract the contribution from a
Standard-Model Higgs of mass $M_0$, since this is already taken into account
in the fits, and find:
\begin{align} \label{Eq:deltarho-HG}
&A_{WW}^{HG}(0)-\cos^2\theta_W\, A_{ZZ}^{HG}(0)\nonumber\\
&=\frac{g^2}{64\pi^2}\Bigl[\sum_j
[\cos\beta\, R_{j1}+\sin\beta\, R_{j2}]^2 \,
\Bigl(3F_{\Delta\rho}(M_Z^2,M_j^2)-3F_{\Delta\rho}(M_W^2,M_j^2)\Bigl) 
\nonumber \\
&\quad
+3F_{\Delta\rho}(M_W^2,M_0^2)
-3F_{\Delta\rho}(M_Z^2,M_0^2)\Bigr].
\end{align}
From the electroweak fits, we take $M_0=129~\text{GeV}$, but note that this
value is not very precise \cite{:2005em}.  In the CP-conserving limit, these
expressions (\ref{Eq:deltarho-HH}) and (\ref{Eq:deltarho-HG}) simplify
considerably, since terms with $R_{13}$, $R_{23}$, $R_{31}$, and $R_{32}$ are
absent.

In order to keep these additional contributions (\ref{Eq:deltarho-HH}) and
(\ref{Eq:deltarho-HG}) small, the charged Higgs boson should not be coupled
too strongly to the $W$ if its mass is far from those of its neutral partners.
As a measure of the ``tolerance'' we take $3\sigma$, i.e., we impose
$|\Delta\rho|\le0.003$.

\subsection{The muon anomalous magnetic moment} 
\label{sect:(g-2)_mu}

The dominant contribution of the Higgs fields to the muon anomalous magnetic
moment, is according to refs.~\cite{Chang:2000ii} and \cite{Cheung:2003pw} due
to the two-loop Barr--Zee effect \cite{Barr:1990vd}, with a photon and a Higgs
field connected to a heavy fermion loop. The contributions are given by
\cite{Cheung:2003pw} in terms of scalar and pseudoscalar Yukawa
couplings. Re-expressed in terms of the Yukawa couplings of (\ref{Eq:H_itt}),
assuming that the muon couples to the Higgs fields like a down quark, i.e., to
$\Phi_1$, we find for the top quark contribution:
\begin{equation} \label{Eq:Delta-amu}
\Delta a_\mu=\frac{N_c \alpha_\text{e.m.}}{4\pi^3 v^2}\,
m_\mu^2 Q_t^2
\sum_j\biggl[R_{j3}^2\, g\biggl(\frac{m_t^2}{M_j^2}\biggr)
-\frac{1}{\cos\beta\sin\beta}\, R_{j1}R_{j2}\, 
f\biggl(\frac{m_t^2}{M_j^2}\biggr)\biggr],
\end{equation}
with $N_c=3$ the number of colours associated with the fermion loop,
$\alpha_\text{e.m.}$ the electromagnetic finestructure constant, $Q_t=2/3$ and
$m_t$ the top quark charge and mass, and $m_\mu$ the muon mass.  The functions
$f$ and $g$ are given in \cite{Barr:1990vd}.  It is worth noting that the
$\tan\beta$ factor associated with the pseudoscalar Yukawa coupling of the
muon is cancelled by an opposite factor associated with the top quark.  While
the first term gives a positive contribution, the second one may have either
sign.

The contribution of the $b$ quark can be obtained from (\ref{Eq:Delta-amu})
by trivial substitutions for $Q_t$ and $m_t$ accompanied by
\begin{equation} \label{Eq:Delta-amu-b}
R_{j3}^2\to\tan^2\beta R_{j3}^2, \quad\text{and}\quad
\frac{1}{\cos\beta\sin\beta}R_{j1}R_{j2}\to \frac{1}{\cos^2\beta}R_{j1}^2
\end{equation}
in the square bracket.

Earlier studies (see, for example, \cite{Chang:2000ii,Cheung:2003pw}) have
focused on the contributions from rather light pseudoscalars and large
$\tan\beta$, where the $b$ and $\tau$ contributions are enhanced by the
substitutions (\ref{Eq:Delta-amu-b}). At high $\tan\beta$, the $b$-quark loop
will indeed dominate. For small values of $\tan\beta$, as are considered here,
the $b$-quark contribution is completely negligible.

The experimental situation is somewhat unclear, depending on how the hadronic
corrections to the running of $\alpha_\text{e.m.}$ are evaluated. The
deviation from the Standard Model can be summarized as \cite{deJong:2005mk}
\begin{equation} \label{Eq:Delta-amu-exp}
\Delta a_\mu^\text{exp}-\Delta a_\mu^\text{SM}
=\begin{cases}
(221-245)\times10^{-11}, \quad &e^+e^-, \\
(62-142)\times10^{-11},  \quad &\tau^+\tau^-, \{e^+e^-\ \&\ \tau^+\tau^-\},
\end{cases}
\end{equation}
and represent 0.7 to 2.8 standard deviations with respect to the data
\cite{Bennett:2004pv}. Two distinct attitudes are here possible. One may
either fit this (positive) deviation with some new physics effect
\cite{Cheung:2003pw}, or one may restrict new physics contributions not to
exceed this contribution (\ref{Eq:Delta-amu-exp}).  We shall here follow this
latter approach, and require the 2HDM contribution (\ref{Eq:Delta-amu}) to be
less than $3\sigma$, i.e., $|\Delta a_\mu|<300\times10^{-11}$.  For the
parameters considered here, $\tan\beta\le{\cal O}(2)$, the 2HDM contribution
to $\Delta a_\mu$ is at most $(\text{a few})\times10^{-11}$ and therefore
plays no role in constraining the model.

\subsection{$R_b$} 
\label{sect:Rb}

The one-loop contributions to the $Zb\bar b$ coupling influence the relative
branching ratio of $Z\to b\bar b$, given by $R_b$, which is known to
0.05\%, or 1.25~MeV precision \cite{Eidelman:2004wy}.  
In the SM there are significant contributions
proportional to $m_t^2$.  In the 2HDM there are additional one-loop
contributions due to triangle diagrams involving charged and (non-standard)
neutral-Higgs fields.  For the CP-conserving case, these were given in
\cite{Denner:1991ie}.  In the general CP-violating case, the charged-Higgs
contribution, Eq.~(4.2) of \cite{Denner:1991ie}, remains unchanged, but we
find that the neutral-Higgs part, Eq.~(4.4), gets modified to
\begin{align} \label{Eq:Rb-Denner-mod}
\delta\Gamma_V^f(H)
&=\frac{\alpha_\text{e.m}^2 N_c\, m_Z}{96\pi \sin^4\theta_W\cos^2\theta_W}\,
\frac{m_b^2}{m_W^2}
\biggl\{\sum_{j=1}^3\biggl[\sum_{k=1}^3
[(\sin\beta R_{j1}-\cos\beta R_{j2})R_{k3} \nonumber \\
& \quad
-(\sin\beta R_{k1}-\cos\beta R_{k2})R_{j3}]
\frac{\tan\beta}{\cos\beta}\,R_{j1}R_{k3}\,
\rho_4(m_Z^2,M_j^2,M_k^2,0) \nonumber \\
& \quad
-(\cos\beta R_{j1}+\sin\beta R_{j2})\frac{R_{j1}}{\cos\beta}\,
\rho_4(m_Z^2,M_j^2,m_Z^2,0) \nonumber \\
& \quad
+2Q_b\sin^2\theta_W(1+2Q_b\sin^2\theta_W)
\biggl(\frac{R_{j1}^2}{\cos^2\beta}+\tan^2\beta R_{j3}^2\biggr)
\rho_3(m_Z^2,M_j^2,0)\biggr] \nonumber \\ 
& \quad
+ 2Q_b\sin^2\theta_W(1+2Q_b\sin^2\theta_W) \,
\rho_3(m_Z^2,m_Z^2,0) \biggr\}.
\end{align}
The functions $\rho_3$ and $\rho_4$ are various combinations of three-point
and two-point loop integrals \cite{Denner:1991ie}.  For the numerical studies,
we use the {\tt LoopTools} package \cite{Hahn:1998yk,vanOldenborgh:1989wn}.
Again, this expression (\ref{Eq:Rb-Denner-mod}) is more complicated than those
of the CP-conserving limit, but the additional terms have little quantitative
importance at low $\tan\beta$.

\begin{table}[ht]
\refstepcounter{table}
\label{table:R_b}
\addtocounter{table}{-1}
\begin{center}
\renewcommand{\tabcolsep}{.75em}
\begin{tabular}{|c|c|c|c|c|}
\hline 
$\tan\beta$& 0.5 & 1.0 & 2.0 \\
\hline
Set B: $M_{H^\pm}=300~\text{GeV}$ & 5.6 & 1.4 & 0.35 \\
Set A: $M_{H^\pm}=500~\text{GeV}$ & 3.3 & 0.83 & 0.21 \\
\hline
\end{tabular} 
\caption{2HDM contribution to $R_b$ for parameter sets A and B,
in units of the uncertainty, $\sigma(R_b)=1.25~\text{MeV}$.}
\end{center}
\end{table}

At small $\tan\beta$, the charged-Higgs contribution, which behaves like
$(m_t/\tan\beta)^2$ due to the $H^+b\bar t$ and $H^-\bar bt$ couplings,
dominates.  For $M_{H^\pm}=500~\text{GeV}$ (Set~A) and
$M_{H^\pm}=300~\text{GeV}$ (Set~B) the 2HDM contributions to $R_b$ are given
in Table~\ref{table:R_b}.  For the two larger values of $\tan\beta$, this is
compatible with the experimental uncertainty, whereas for $\tan\beta=0.5$ it
amounts to a substantial conflict, especially for the lower value of
$M_{H^\pm}$ (Set~B).  The neutral-Higgs contribution, as given by
(\ref{Eq:Rb-Denner-mod}), is smaller, by three orders of magnitude.

\section{Overview over allowed parameters}
\label{sect:parameter-overview}
\setcounter{equation}{0}

\subsection{Variations of mass parameters around Set~A} 
\label{sect:mass-parameters-setA}

We start this discussion of model parameters by a survey of how the allowed
regions of the $\alpha_2$--$\alpha_3$ space depend on the mass parameters, in
particular $M_3$, $M_{H^\pm}$ and $\mu^2$.  It turns out that while stability
is readily satisfied for ``relevant'' mass parameters, unitarity excludes
sizable regions of parameter space. Ignoring experimental constraints, a
low-mass spectrum is in general easier to accommodate than one where some
Higgs particles are heavy. In many cases, non-zero values of $\lambda_6$ and
$\lambda_7$ also have a tendency to reduce the allowed parameter space.

\begin{figure}[htb]
\refstepcounter{figure}
\label{Fig:A-alphas-pos}
\addtocounter{figure}{-1}
\begin{center}
\setlength{\unitlength}{1cm}
\begin{picture}(15.0,11.8)
\put(1.5,-0.7)
{\mbox{\epsfysize=13cm
 \epsffile{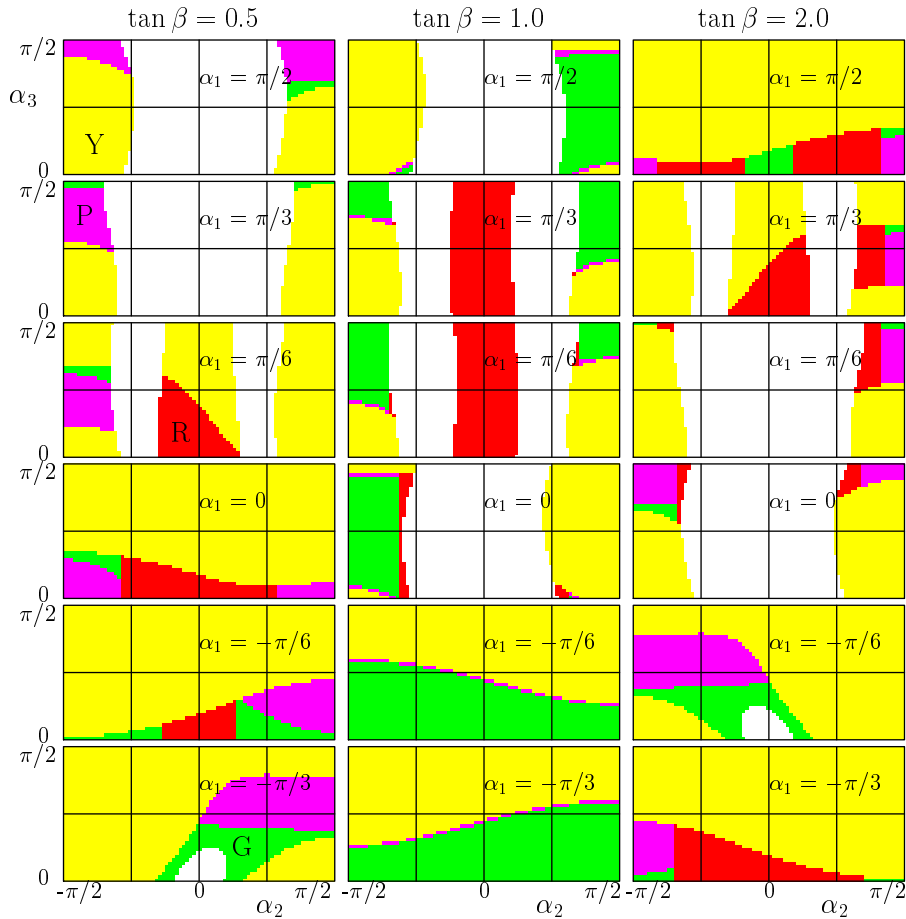}}}
\end{picture}
\caption{Allowed and forbidden regions in the $\alpha_2$--$\alpha_3$ plane,
for $\tan\beta=0.5$, 1 and 2, parameter Set~A and selected values of
$\alpha_1$.  Green (G): stability, unitarity, direct search and $\Delta\rho$
constraints satisfied; red (R) [purple (P)]: stability, unitarity satisfied,
but direct search [and $\Delta\rho$] constraints not satisfied; yellow (Y):
stability (but not unitarity) satisfied; white: stability violated.}
\end{center}
\end{figure}

In Fig.~\ref{Fig:A-alphas-pos} we show for Set~A the allowed regions in the
$\alpha_2$--$\alpha_3$ plane, for a few representative values of $\alpha_1$,
focusing on stability (or positivity) and unitarity.  For the considered
parameters, much of the $\alpha_2$--$\alpha_3$ plane is actually in violation
of stability, as shown in white.  Next, there are significant areas (yellow)
where stability is satisfied, but (tree-level) unitarity is violated.
Finally, in darker colours, we indicate where also unitarity holds.  Part of
these areas (typically, small $|\alpha_2|$, as discussed in
sect.~\ref{sect:LEP-noHiggs}) are in conflict with the direct search (LEP)
data, and indicated in red.  Some areas (purple, typically, larger
$|\alpha_2|$) violate the $\Delta\rho$ constraint, whereas the remaining areas
in green give viable models.  The symmetry given by
Eq.~(\ref{Eq:tanbeta-symm}) is evident in the figure: the case of
$\tan\beta=2$ can be obtained from the case of $\tan\beta=0.5$ by suitable
reflections, apart from the experimental constraints, which in the 2HDM (II)
distinguish $\tan\beta$ and $\cot\beta$.  For $\tan\beta>1$, the direct search
constraints shift from those of (\ref{Eq:LEP-smalltanbeta}) towards those of
(\ref{Eq:LEP-largetanbeta}).

While this figure can not be directly compared with Fig.~7 of
ref.~\cite{Khater:2003wq}, since we here keep $M_3$ fixed, it was found that
the unitarity constraints of \cite{Akeroyd:2000wc,Ginzburg:2003fe} are more
restrictive than the order-of-magnitude estimate adopted in
\cite{Khater:2003wq}.
It should also be noted that with $\Im\lambda_5=0$, as given by the
parameter Set~A, $\Im\lambda_6$ and $\Im\lambda_7$ will be non-zero
in the general CP-violating case.

\begin{figure}[hbt]
\refstepcounter{figure}
\label{Fig:alphas-three-mh3-mhch}
\addtocounter{figure}{-1}
\begin{center}
\setlength{\unitlength}{1cm}
\begin{picture}(15.0,11.8)
\put(-1.2,-0.9)
{\mbox{\epsfysize=13cm
 \epsffile{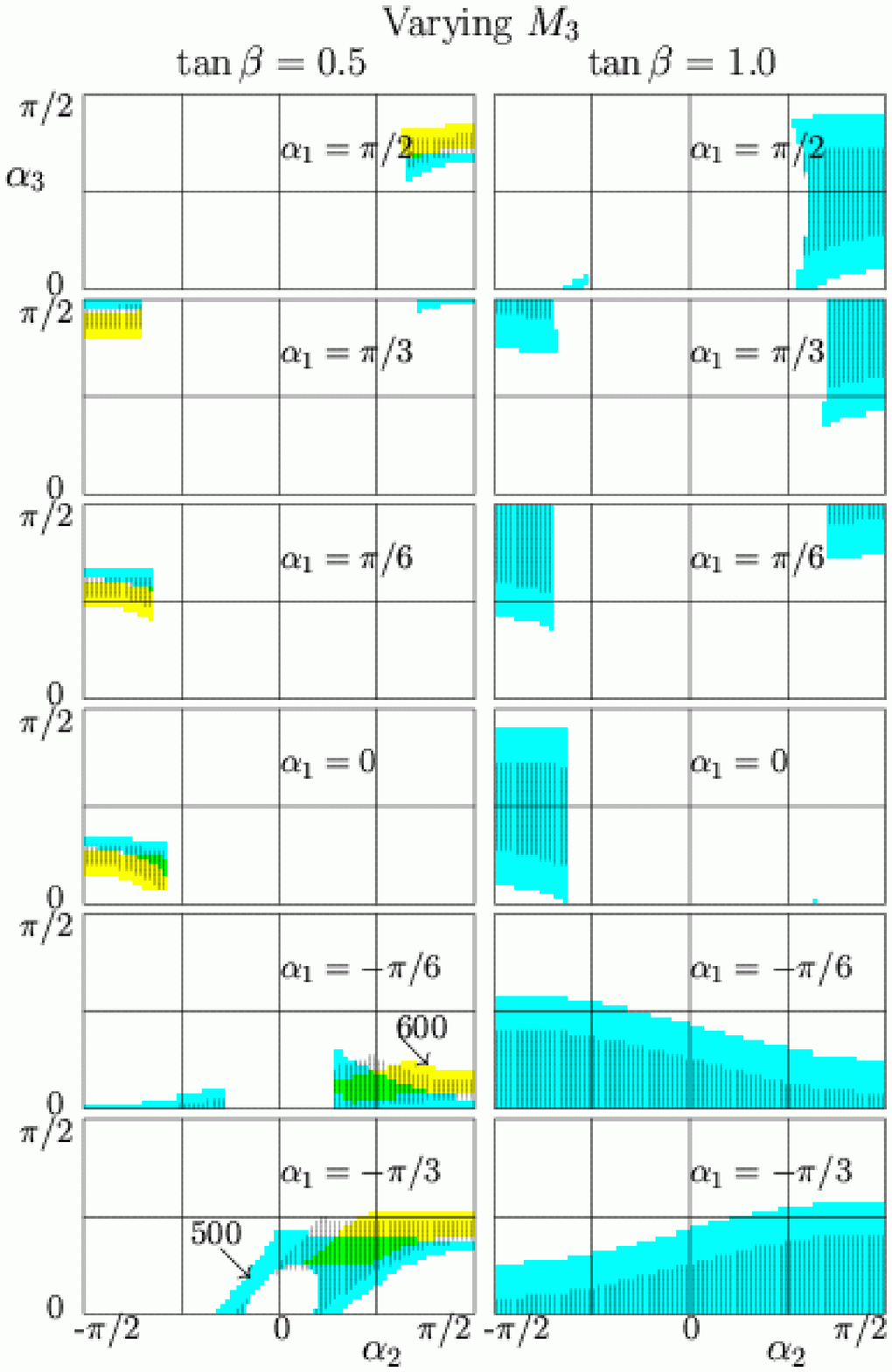}}
 \mbox{\epsfysize=13cm
 \epsffile{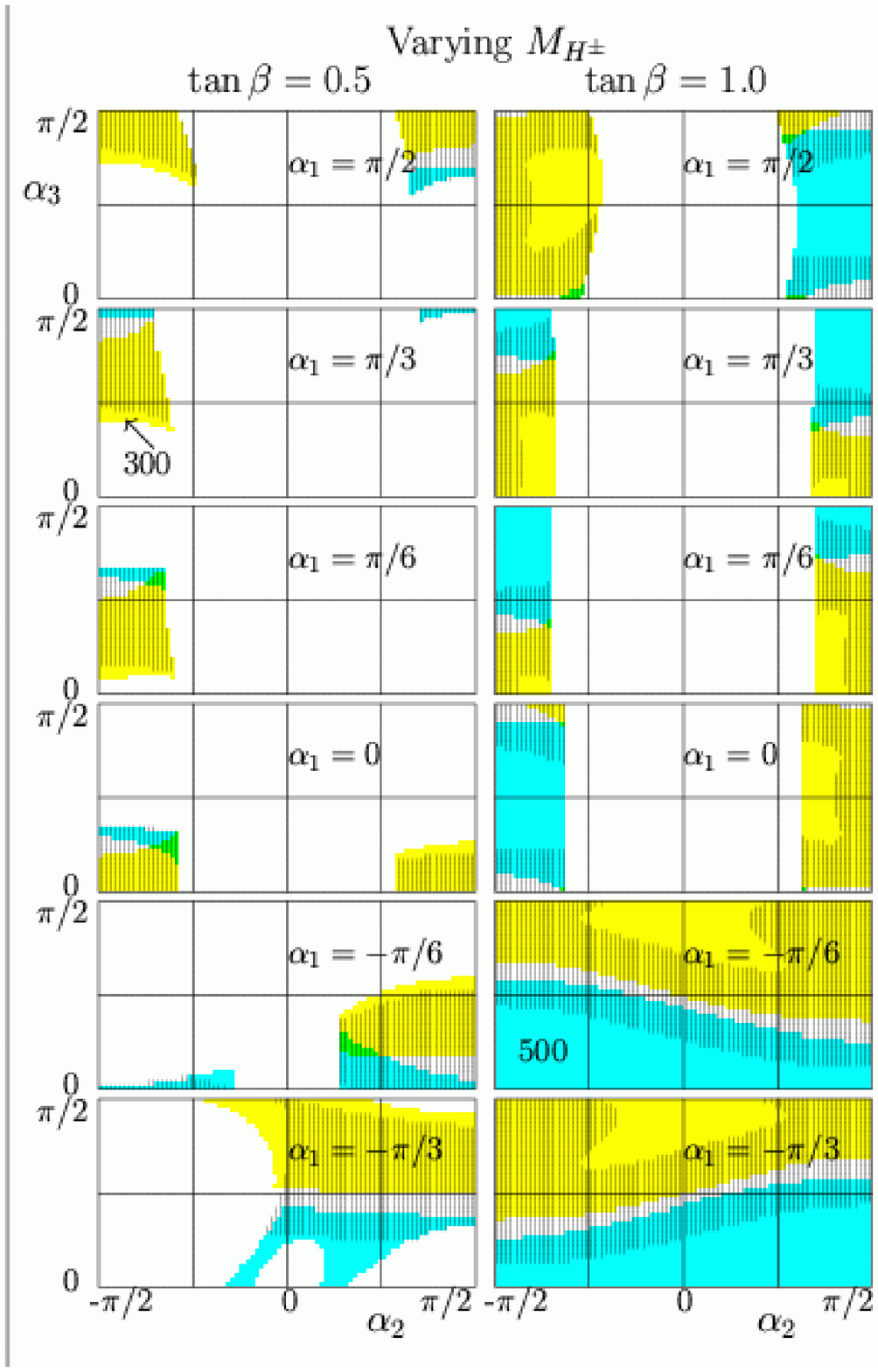}}}
\end{picture}
\vspace*{2mm}
\caption{Physically allowed regions in the $\alpha_2$--$\alpha_3$ plane, for
$\tan\beta=0.5$ and 1, and selected values of $\alpha_1$.  
Left: Variations of $M_3$ around parameter Set~A. Blue: $M_3=500~\text{GeV}$
(Set~A), vertical lines: $M_3=550~\text{GeV}$, 
yellow: $M_3=600~\text{GeV}$.
Note that some regions overlap. In particular,
green denotes regions allowed for both $M_3=500~\text{GeV}$ and
$M_3=600~\text{GeV}$.
Right: Variations of $M_{H^\pm}$ around parameter Set~A.
Yellow: $M_{H^\pm}=300~\text{GeV}$,
vertical lines: $M_{H^\pm}=400~\text{GeV}$,
blue: $M_{H^\pm}=500~\text{GeV}$ (Set~A).}
\end{center}
\vspace*{-4mm}
\end{figure}

\paragraph{Dependence on $M_3$ and $M_2$.}
The dependence of the allowed regions on $M_3$, the heaviest neutral Higgs
boson, is illustrated in Fig.~\ref{Fig:alphas-three-mh3-mhch} (left), for the
other parameters kept fixed at the values of parameter Set~A.  The figure
shows the allowed regions for $M_3=500$~GeV (blue, default of
Fig.~\ref{Fig:A-alphas-pos}), 550~GeV (vertical lines) and 600~GeV (yellow).
Smaller allowed regions are also found at 450 and 650~GeV (not shown), but
nothing neither at 400~GeV nor at 700~GeV.

As $M_2$ approaches $M_1$, there are still regions where stability is
satisfied, but unitarity is only satisfied in very small regions.  Similarly,
as discussed above, when $M_2$ approaches $M_3$, the allowed regions tend to
be restricted to small values of $|\alpha_2|\to0$.  Thus, by
(\ref{Eq:altilde-alb}), there is in this limit of $M_1\ll M_2\lsim M_3$,
little CP violation in the top-quark sector.

\paragraph{Dependence on $M_H^\pm$.}
The dependence of the allowed regions on $M_{H^\pm}$, the charged Higgs boson,
is illustrated in Fig.~\ref{Fig:alphas-three-mh3-mhch} (right), for the other
parameters kept fixed at the values of parameter Set~A. The allowed region is
seen to increase a bit for lower values of $M_{H^\pm}$, but such values are
constrained by the $\Delta M_{B_d}$, $R_b$ \cite{Eidelman:2004wy} and $b\to
s\gamma$ data \cite{Gambino:2001ew}, especially at low values of $\tan\beta$.
On the other hand, the allowed regions shrink for high values of $M_{H^\pm}$.
Nothing is allowed at $M_H^\pm\ge 600$~GeV.  The shrinking of the allowed
regions for high values of $M_{H^\pm}$ is due to the $\Delta\rho$ constraint
at large $|\alpha_2|$ and/or negative $\alpha_1$, as well as the direct search
(LEP) constraint at small $|\alpha_2|$. Eventually (like for high values of
$M_3$) also the unitarity constraint excludes high values of $M_H^\pm$.

\begin{figure}[htb]
\refstepcounter{figure}
\label{Fig:alphas-musq}
\addtocounter{figure}{-1}
\begin{center}
\setlength{\unitlength}{1cm}
\begin{picture}(15.0,11.5)
\put(-1.2,-0.9)
{\mbox{\epsfysize=13cm
 \epsffile{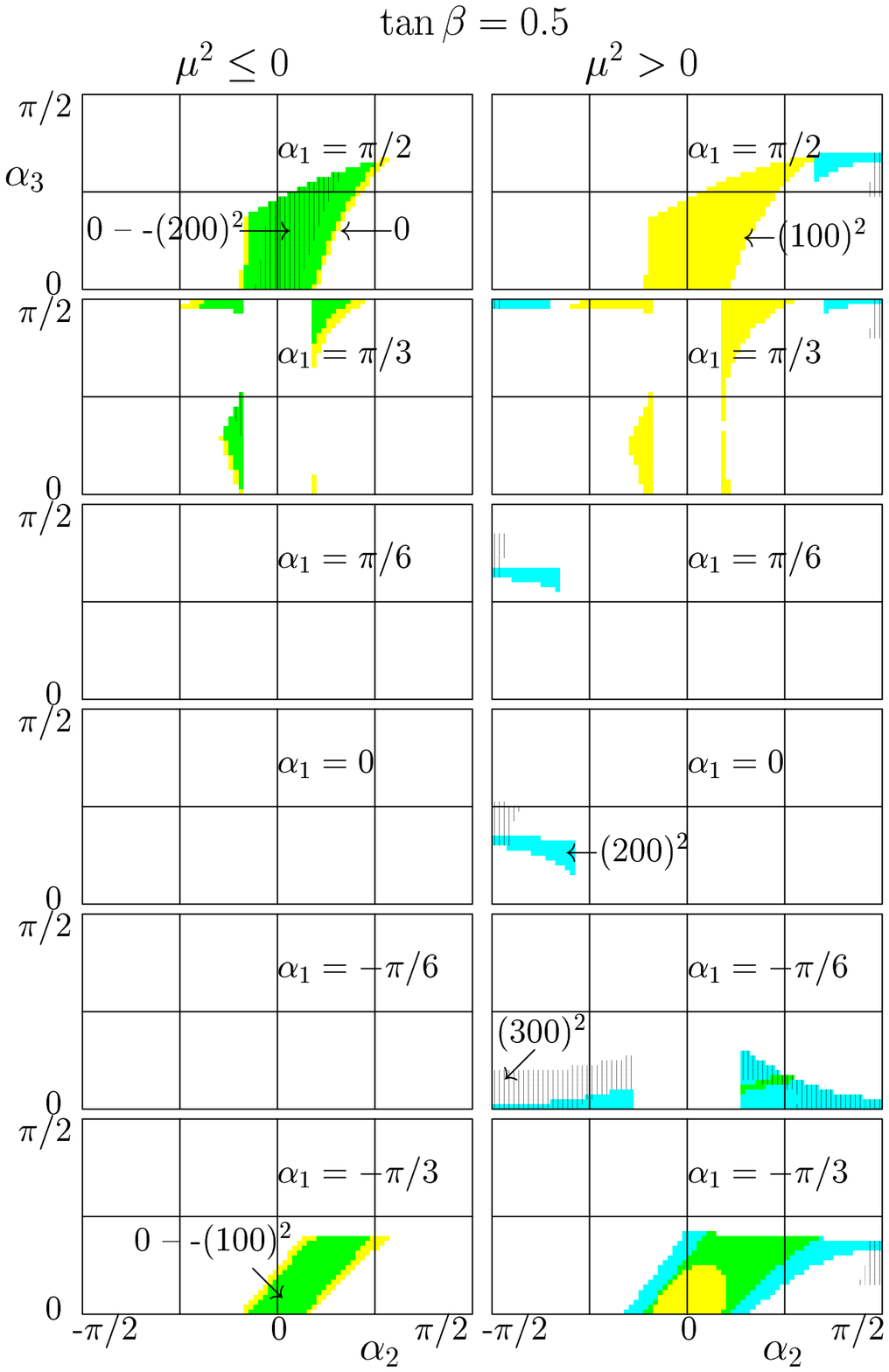}}
 \mbox{\epsfysize=13cm
 \epsffile{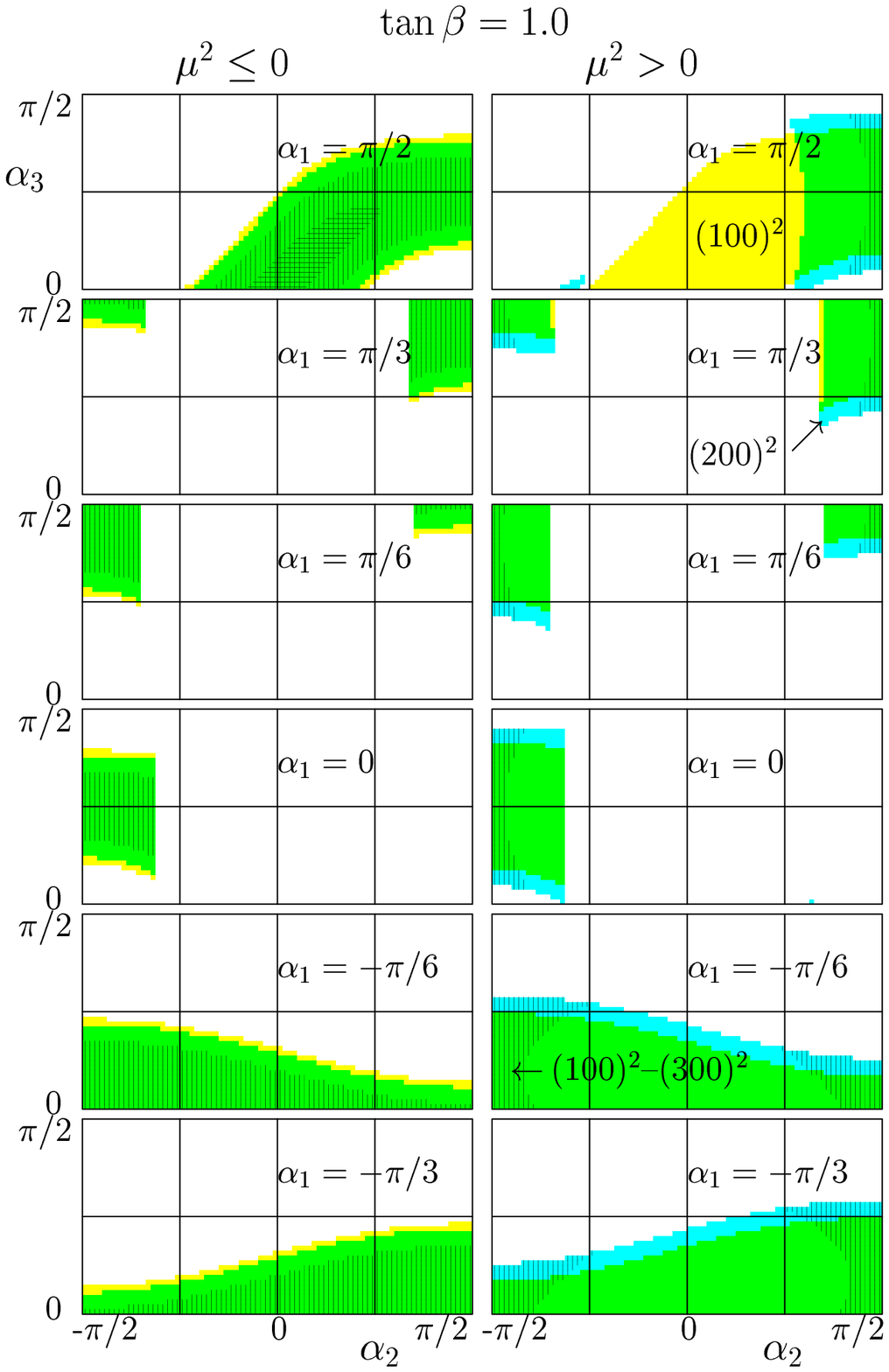}}}
\end{picture}
\vspace*{2mm}
\caption{Physically allowed regions in the $\alpha_2$--$\alpha_3$ plane, for
$\tan\beta=0.5$ and 1, and selected values of $\alpha_1$.
Variations around parameter Set~A; $\mu^2\le0$ and $\mu^2>0$ shown separately.
$\mu^2\le0$:
Yellow: $\mu^2=0$, green: $\mu^2=-(100~\text{GeV})^2$,
vertical lines: $\mu^2=-(200~\text{GeV})^2$.
$\mu^2>0$:
Yellow: $\mu^2=(100~\text{GeV})^2$,
blue: $\mu^2=(200~\text{GeV})^2$, vertical lines: $\mu^2=(300~\text{GeV})^2$.
Note that some regions overlap. In particular, for $\mu^2>0$, green is allowed
for both $\mu^2=(100~\text{GeV})^2$ and $\mu^2=(200~\text{GeV})^2$.}
\end{center}
\vspace*{-4mm}
\end{figure}

\paragraph{Dependence on $\mu^2$.}
Since $\lambda_4+\Re\lambda_5$ is bounded from below by the stability
requirement [see (\ref{Eq:lambda45-pos})], the parameter $\mu^2$ normally
plays a role in pushing $M_{H^\pm}$ to values that are high enough to evade
the $R_b$, $\Delta M_{B_d}$ \cite{Eidelman:2004wy} and $b\to s\gamma$
constraints.  In Fig.~\ref{Fig:alphas-musq} we show how the allowed region in
the $\alpha_2$--$\alpha_3$ plane shrinks for ``low'' and ``high'' values of
$\mu^2$, when the Higgs boson masses and other parameters are kept fixed.  For
the considered spectrum of masses, a range of negative values of $\mu^2$ is
allowed (for $\mu^2=-(400~\text{GeV})^2$ there are no allowed regions).  For
increasing positive values of $\mu^2$, the allowed regions shrink away before
$\mu^2=(400~\text{GeV})^2$ (which is not allowed).  Of course, these critical
values depend on the mass spectrum adopted in parameter Set~A.

\subsection{Variations of mass parameters around Set~B} 
\label{sect:mass-parameters-setB}

We shall here briefly review the light-Higgs scenario of parameter Set~B, 
but recall that the $\tan\beta=0.5$ case is essentially ruled out by the 
$\Delta M_b^0$ data.

\begin{figure}[htb]
\refstepcounter{figure}
\label{Fig:B-alphas-pos}
\addtocounter{figure}{-1}
\begin{center}
\setlength{\unitlength}{1cm}
\begin{picture}(15.0,11.5)
\put(1.5,-0.7)
{\mbox{\epsfysize=13cm
 \epsffile{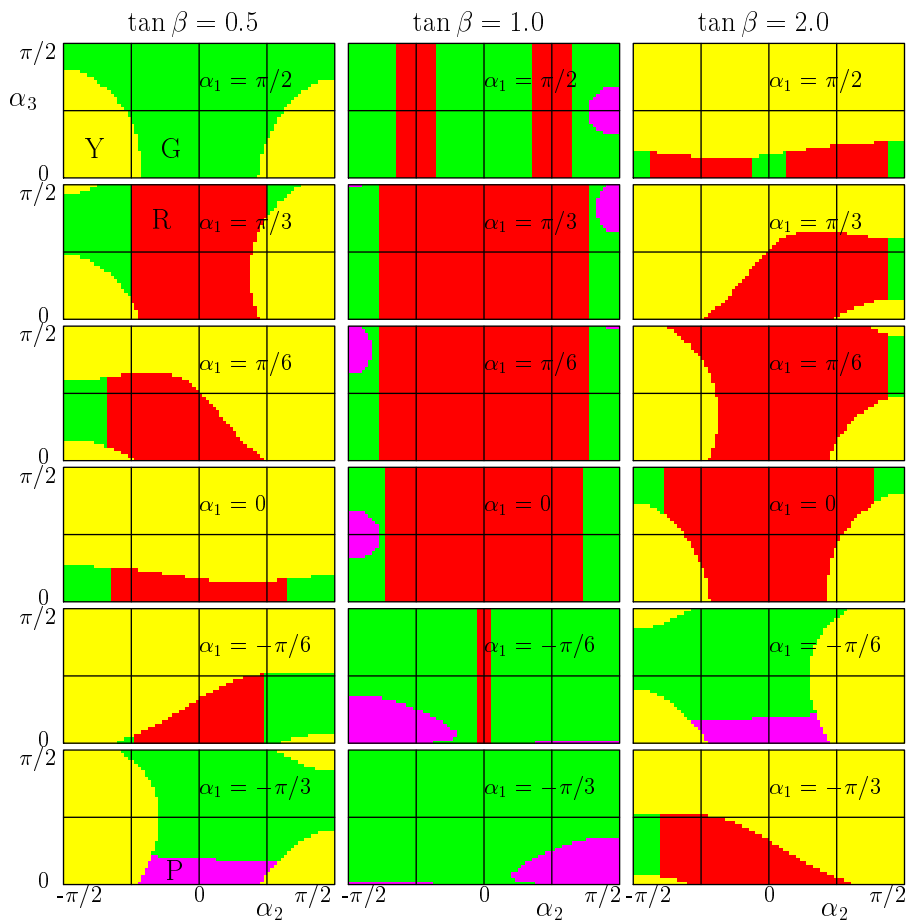}}}
\end{picture}
\caption{Stability and unitarity in the $\alpha_2$--$\alpha_3$ plane, for
$\tan\beta=0.5$, 1 and 2, parameter Set~B  with $M_3=400~\text{GeV}$, $\mu^2=0$
and selected values of $\alpha_1$.
Colour codes as in Fig.~\ref{Fig:A-alphas-pos}.}
\end{center}
\end{figure}

In Fig.~\ref{Fig:B-alphas-pos} we show for $\mu^2=0$ how stability can be
satisfied in the {\it whole} $\alpha_2$--$\alpha_3$ plane. For $\tan\beta=1$,
stability and unitarity would actually allow any value of the rotation matrix
$R$, i.e., any values of $\alpha_1$, $\alpha_2$ and $\alpha_3$.  However,
except for this special case, unitarity is only satisfied in parts of the
plane.

The direct search (LEP) constraint severely cuts into the allowed regions of
this light-Higgs (set~B) scenario, with $C^2=0.1$ \cite{Boonekamp:2004ae}, see
sect.~\ref{sect:LEP-noHiggs}.  As noted there, this constraint tends to
exclude the central range of small $|\alpha_2|$ and positive and small
$\alpha_1$. 

\begin{figure}[htb]
\refstepcounter{figure}
\label{Fig:alphas-setB-M3=400-musq=40000}
\addtocounter{figure}{-1}
\begin{center}
\setlength{\unitlength}{1cm}
\begin{picture}(15.0,11.5)
\put(1.5,-0.7)
{\mbox{\epsfysize=13cm
 \epsffile{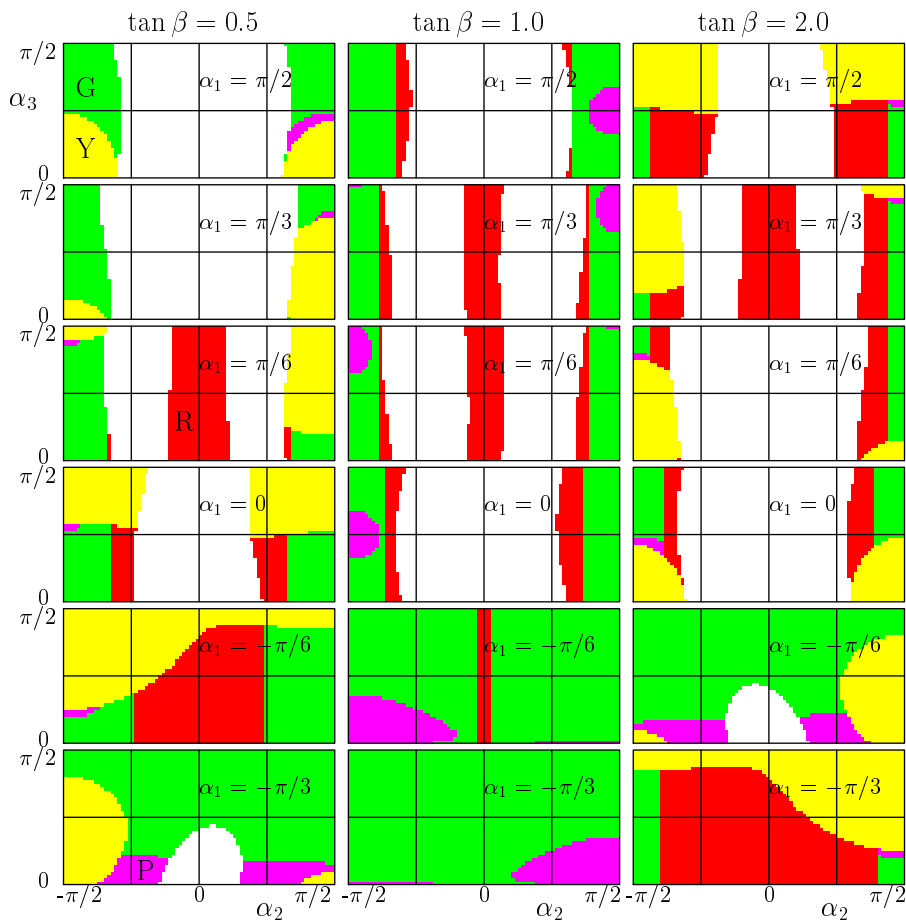}}}
\end{picture}
\caption{Stability and unitarity in the $\alpha_2$--$\alpha_3$ plane, for
$\tan\beta=0.5$, 1 and 2, parameter Set~B with $M_3=400~\text{GeV}$,
$\mu^2=(200~\text{GeV})^2$ and selected values of $\alpha_1$.  
Colour codes as in Fig.~\ref{Fig:A-alphas-pos}.}
\end{center}
\end{figure}

Increasing $\mu^2$ to $(200~\text{GeV})^2$, the most striking change is
perhaps the emergence of large regions where stability is not satisfied
(indicated in white in Fig.~\ref{Fig:alphas-setB-M3=400-musq=40000}).
Another interesting observation is that for $\tan\beta=1.0$ and negative 
values of $\alpha_1$, the picture is little changed from the
case of $\mu^2=0$.

With a higher value of $M_3$ ($M_3=600$~GeV, but $\mu^2=0$), the unitarity
constraint excludes most of the $\alpha_2$--$\alpha_3$ plane.  
For $\tan\beta=1.0$ the whole plane is excluded.

\subsection{Non-zero values of $\Im\lambda_5$, $\Re\lambda_6$, $\Re\lambda_7$}

This subsection will present a brief discussion of how the allowed regions get
modified for non-zero values of parameters which are normally set to zero, in
order to control the amount of Flavour-Changing Neutral Currents.  Thus,
adopting a non-zero value for any of them in a ``realistic'' model would have
to be done with an eye to these effects.

\subsubsection{Non-zero values of $\Im\lambda_5$}

The cases of positive and negative values of $\Im\lambda_5$ can be related.
This is seen as follows: by the transformation C3 of (\ref{Eq:symm-C}) and
(\ref{Eq:symm-C-M}), the mass-squared elements ${\cal M}^2_{13}$ and ${\cal
M}^2_{23}$ flip signs, and the auxiliary quantities $\Im\lambda_6$ and
$\Im\lambda_7$ change signs without altering the stability or unitarity
constraints. The only effect will be that CP-violating effects change sign.
Thus, we may restrict the discussion of non-zero $\Im\lambda_5$ to
$\Im\lambda_5>0$. However, the full range (\ref{Eq:angular-range}) of
$\alpha_3$ now has to be considered.  

We show in Fig.~\ref{Fig:alphas-lambda5_im}, for some values of $\tan\beta$
and $\alpha_1$, and for parameter Set~A how non-zero values of $\Im\lambda_5$
also provide allowed regions in the $\alpha_2$--$\alpha_3$ plane.  But these
are typically smaller for higher values of $\Im\lambda_5$, and rather
scattered. However, they can lead to significant CP violation, since allowed
ranges of $|\alpha_2|$ tend to be at intermediate values of $|\alpha_2|$ [see
Eq.~(\ref{Eq:CPV-condition})].

\begin{figure}[htb]
\refstepcounter{figure}
\label{Fig:alphas-lambda5_im}
\addtocounter{figure}{-1}
\begin{center}
\setlength{\unitlength}{1cm}
\begin{picture}(15.0,13.0)
\put(3.0,-0.7)
{\mbox{\epsfysize=14cm
 \epsffile{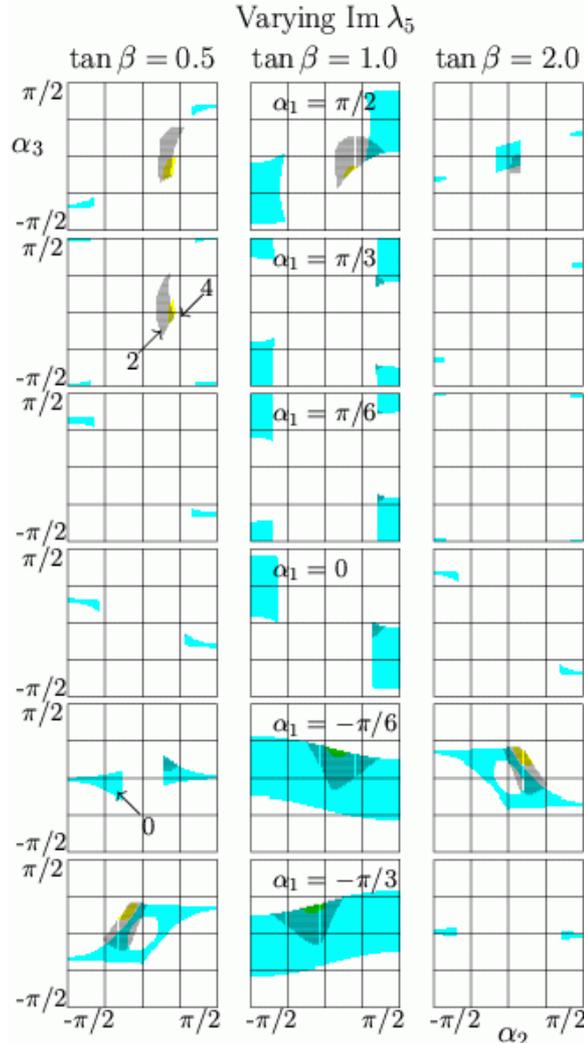}}}
\end{picture}
\caption{Physically allowed regions in the $\alpha_2$--$\alpha_3$ plane, for
$\tan\beta=0.5$, 1 and 2, and selected values of $\alpha_1$.  
Variations of $\Im\lambda_5$ around parameter Set~A. Blue: $\Im\lambda_5=0$,
vertical lines: $\Im\lambda_5=2$, yellow: $\Im\lambda_5=4$.}
\end{center}
\end{figure}

\subsubsection{Non-zero values of $\Re\lambda_6$ or $\Re\lambda_7$}

Up to this point, we have kept $\Re\lambda_6=\Re\lambda_7=0$.  However, we
have treated $\Im\lambda_6$ and $\Im\lambda_7$ as auxiliary quantities derived
from the spectrum, the rotation matrix and $\Im\lambda_5$ via
Eq.~(\ref{Eq:Im-lam67}). In general, they will be non-zero.  This might lead
to too large flavour-changing neutral couplings, due to the violation of the
$Z_2$ symmetry \cite{Weinberg:1976hu,Glashow:1976nt,Branco}. We have not
investigated this constraint quantitatively, but note that in many cases the
imaginary parts of $\lambda_6$ and $\lambda_7$ can be shifted to the imaginary
part of $\lambda_5$ [see (\ref{Eq:M_ij})]. This will however lead to a
modification of the rotation matrix $R$ and/or the spectrum, by for example a
shift in $M_3$ according to the approach of \cite{Khater:2003wq}.

\begin{figure}[htb]
\refstepcounter{figure}
\label{Fig:alphas-lambda6_re=min1}
\addtocounter{figure}{-1}
\begin{center}
\setlength{\unitlength}{1cm}
\begin{picture}(15.0,11.8)
\put(1.5,-0.7)
{\mbox{\epsfysize=13cm
 \epsffile{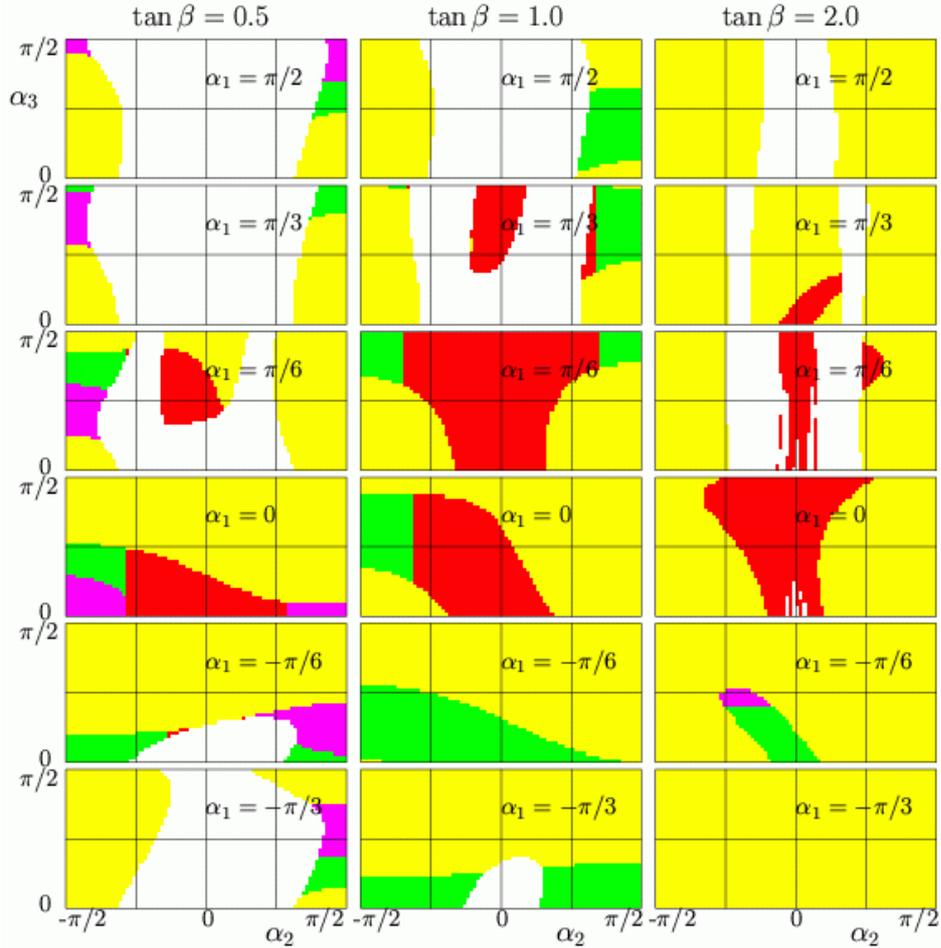}}}
\end{picture}
\vspace*{2mm}
\caption{Physically allowed regions in the $\alpha_2$--$\alpha_3$ plane, for
$\tan\beta=0.5$, 1 and 2, and selected values of $\alpha_1$.
Parameters correspond to Set~A, except that $\Re\lambda_6=-1$.
Colour codes as in Fig.~\ref{Fig:A-alphas-pos}.}
\end{center}
\vspace*{-4mm}
\end{figure}

The Yukawa interactions (see sect.~\ref{sect:Yukawa}) couple the Higgs fields
to a left-handed doublet and a right-handed singlet quark field. However,
these need not be in the flavour basis in which the mass matrices are
diagonal. The $Z_2$ symmetry, which is imposed to stabilize Model~II
\cite{Weinberg:1976hu,Glashow:1976nt} is broken by the $m_{12}^2$ and
$\Im\lambda_5$ terms, as well as by the $\lambda_6$ and $\lambda_7$ terms.
However, one may adopt the attitude that these terms, which arise naturally
in the MSSM
\cite{Wess:1974tw,Fayet:1976cr,Dimopoulos:1981zb,Nilles:1983ge,Haber:1984rc}
are constrained and subdominant.  Additionally, one may argue that the
FCNC's are suppressed by powers of the quark masses
\cite{Cheng:1987rs,Atwood:1996vj}, thus evading the experimental constraints
involving the first two fermion generations.

We shall here only discuss the case of {\it either} $\Re\lambda_6$ {\it or}
$\Re\lambda_7$ being non-zero, the other being zero.  Because of the symmetry
(\ref{Eq:tanbeta-symm}), (\ref{Eq:tanbeta-symm-part2}), we may restrict this
discussion to $\Re\lambda_6\ne0$, $\Re\lambda_7=0$. Analogous results for
$\Re\lambda_6=0$, $\Re\lambda_7\ne0$ can be obtained from these by the
inversion $\tan\beta\leftrightarrow 1/\tan\beta$ according to
(\ref{Eq:tanbeta-symm}) and (\ref{Eq:tanbeta-symm-part2}).

Even though $\Re\lambda_6$ (or $\Re\lambda_7$) significantly different from
zero may turn out to be ruled out by the constraints on FCNC's, we find it
instructive to see how the otherwise allowed regions change when these
parameters are introduced.

In Fig.~\ref{Fig:alphas-lambda6_re=min1} we show the ``allowed'' regions
corresponding to parameter Set~A, except that $\Re\lambda_6=-1$. The allowed
regions are qualitatively rather similar to those of
Fig.~\ref{Fig:A-alphas-pos}. However, with $\Re\lambda_6=+1$, stability tends
to be violated in most of the $\alpha_2$--$\alpha_3$ plane. Also, when
$\Re\lambda_6$ decreases to $-2$ or $-3$, there is nothing allowed at
$\tan\beta=1$ and $\tan\beta=0.5$, respectively.

\section{CP violation in $t\bar t$ production at the LHC}
\label{sect:ttbar}
\setcounter{equation}{0}
In order to illustrate the CP-violating effects that can be observed
at the LHC, resulting from mixing in the Higgs sector, we consider
the process
\begin{equation} \label{Eq:pp-to-ttbar}
pp \to t \bar t +X, 
\end{equation}
which at high energies is dominated by the underlying process
\begin{equation}
gg \to t \bar t.
\end{equation}
In the presence of CP violation, correlations will then be induced at the
parton level involving the $t$ and $\bar t$ spins, denoted $\boldsymbol{s}_1$
and $\boldsymbol{s}_2$, and their c.m.\ momentum $\boldsymbol{p}$
\cite{Bernreuther:1993hq}:
\begin{equation}      \label{Eq:parton-CPV}
\langle\hat{\boldsymbol{p}}\cdot(\boldsymbol{s}_1-\boldsymbol{s}_2)\rangle,
\quad \text{and}\quad
\langle\hat{\boldsymbol{p}}\cdot(\boldsymbol{s}_1\times\boldsymbol{s}_2)
\rangle.
\end{equation}
These correlations are determined by the CP-violating combination
(\ref{Eq:gamma_CP-i}) of Yukawa couplings, multiplied by certain loop
integrals, and convoluted over the gluon--gluon c.m.\ energy
\cite{Bernreuther:1993hq,Khater:2003wq}.

The $t$ and $\bar t$ quarks decay fast enough that hadronization effects do
not smear out these CP-odd correlations. They can thus be treated
perturbatively. Consider the semileptonic decays:
\begin{equation}
\label{Eq:semileptonic}
t \to l^+ \nu_l b, \quad \bar t \to l^- \bar\nu_l \bar b,
\end{equation}
with $l=e^-$ or $\mu^-$.
The lepton energies will inherit an asymmetry from the first correlation
given in Eq.~(\ref{Eq:parton-CPV}), accessible via the observable:
\begin{equation} \label{Eq:A_1}
A_1=E_+-E_-.
\end{equation}
An important question is whether or not this can be large enough
to be measurable. 

In order to have a significant observation,
the expectation value $\langle A_1\rangle$ must
compare favourably with the statistical fluctuations, which
behave like $\sqrt{N}$, where $N$ is the number of events.
In order to assess this, it is convenient to consider the 
``signal to noise'' ratio \cite{Bernreuther:1993hq},
\begin{equation}  \label{Eq:S/N}
\frac{S}{N}=\frac{\langle A_1\rangle}
{\sqrt{\langle A_1^2\rangle-\langle A_1\rangle^2}},
\end{equation}
where the denominator gives the statistical width of the observable
(\ref{Eq:A_1}).

Other limitations of this approach include the need for very good lepton
energy calibration, and the assumption of no (anomalous) right-handed
couplings in the $tbW$ coupling (see \cite{Kraan:2006jd}).

We display the ratio $S/N$ in Fig.~\ref{Fig:a1-s}, as a function of the mass
of the lightest Higgs boson, $M_1$, for two rotation matrices $R$, given by
the angles $\{\alpha_1,\alpha_2,\alpha_3\}$, and for two values of
$\tan\beta$.  These parameters are chosen such that the model is consistent
(stability and unitarity) and satisfies the experimental constraints (see the
discussion in Sect.~\ref{sect:mass-parameters-setA} and
Fig.~\ref{Fig:A-alphas-pos}).  The actual evaluation of this quantity
(\ref{Eq:S/N}) follows the approach of \cite{Khater:2003wq}, using the {\tt
LoopTools} package for the loop evaluations
\cite{Hahn:1998yk,vanOldenborgh:1989wn} and the {\tt CTEQ6} parton
distribution functions \cite{Pumplin:2002vw} to describe the gluon content of
the proton.

\begin{figure}[htb]
\refstepcounter{figure}
\label{Fig:a1-s}
\addtocounter{figure}{-1}
\begin{center}
\setlength{\unitlength}{1cm}
\begin{picture}(15.0,6.5)
\put(-1.5,-0.9)
{\mbox{\epsfysize=7.5cm
 \epsffile{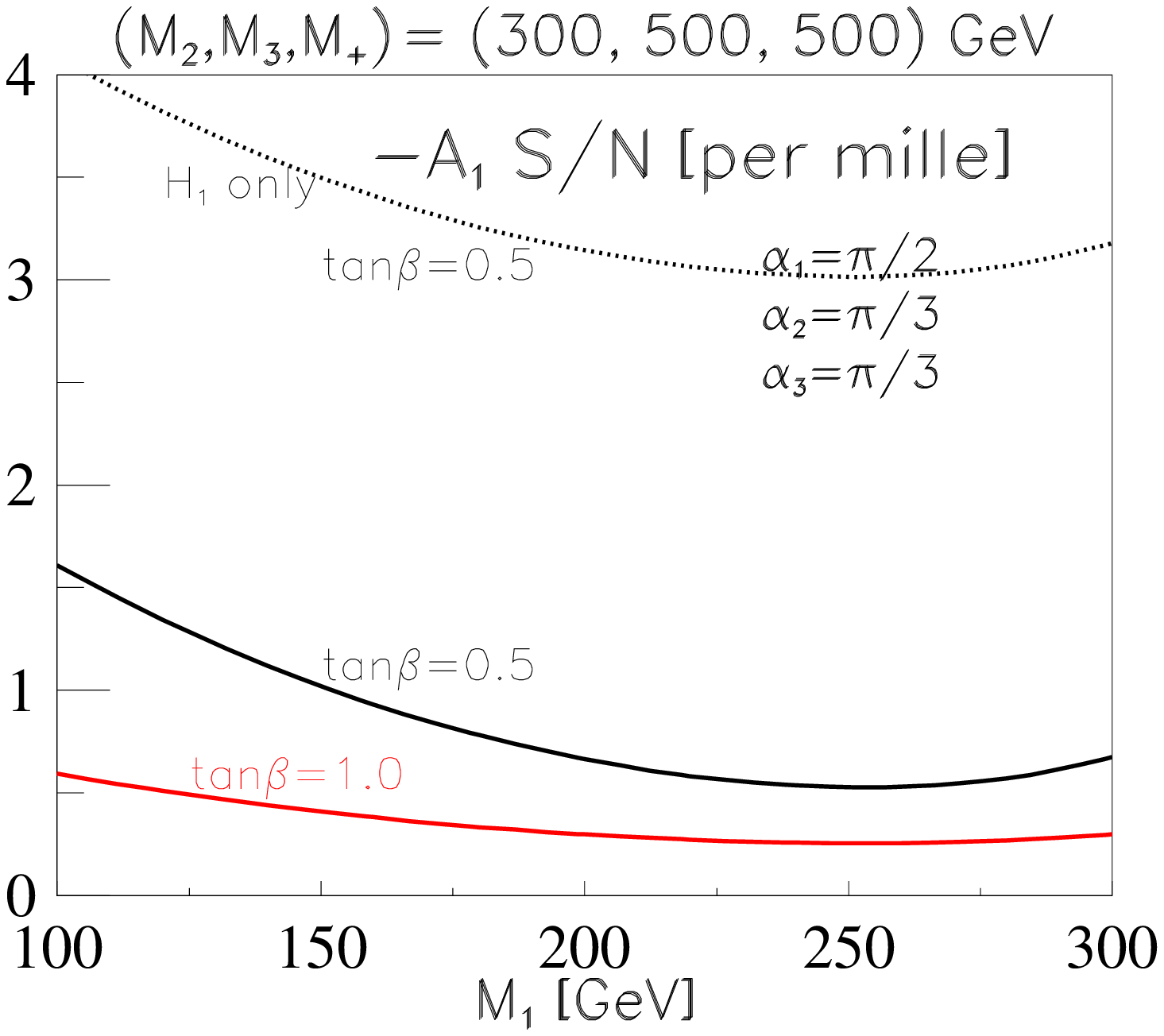}}
 \mbox{\epsfysize=7.5cm
 \epsffile{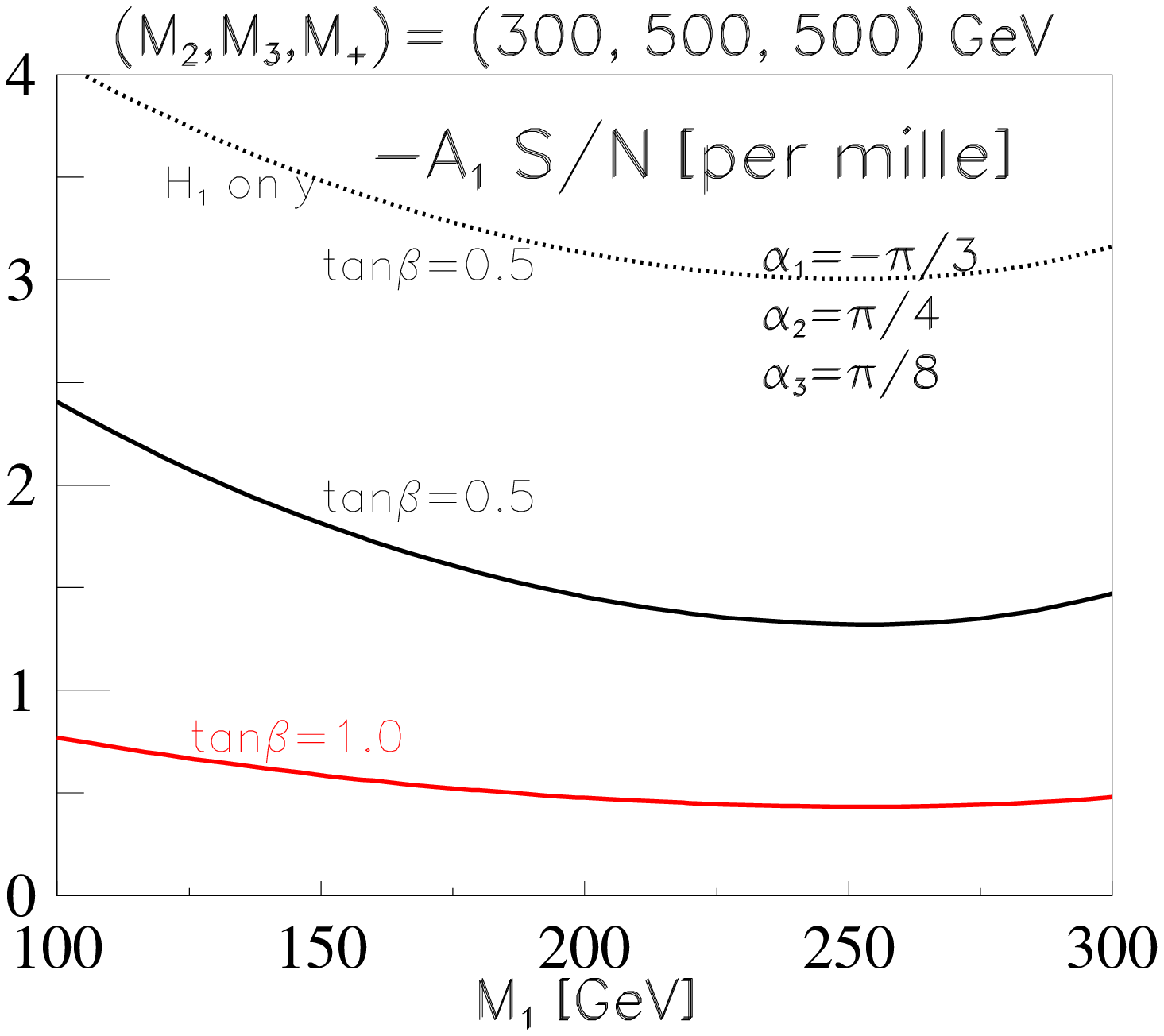}}}
\end{picture}
\vspace*{2mm}
\caption{Signal to noise ratio vs.\ $M_1$, otherwise parameter Set~A.
Rotation matrix $R$ defined by angles $\{\alpha_1,\alpha_2,\alpha_3\}$ as
indicated.  Actually, in the left panel, we show $S/N$ for $-A_1$.
Dotted curves: contribution of the lightest Higgs boson only.}
\end{center}
\vspace*{-4mm}
\end{figure}

While the allowed regions, for $\tan\beta=0.5$, in Fig.~\ref{Fig:A-alphas-pos}
only are rather small and scattered, they remain allowed as $M_1$ is increased
towards $M_2$.  For these two rotation matrices, the contribution of the
lightest Higgs boson, $H_1$, is actually the same (apart from the over-all
sign), as shown by the dotted curves. In terms of the quantities
(\ref{Eq:gamma_CP-i}), $|\gamma_{CP}^{(1)}|=1.936$ is the same for both cases.
However, for the case shown on the left, the contribution of $H_2$ (with
$M_2=300$~GeV), is with $\gamma_{CP}^{(2)}/\gamma_{CP}^{(1)}=-0.75$ more
efficient in reducing the effect of $H_1$ than for the case on the right,
where the corresponding ratio is $-0.44$.

In general, the values fall with increasing $M_1$, since the loop integrals
decrease. However, there is a resonance in one diagram at
$M_j=2m_t\sim350$~GeV (for more details, see \cite{Khater:2003wq}), this is
the reason for the increase beyond 250~GeV.

\begin{figure}[htb]
\refstepcounter{figure}
\label{Fig:a1-s-vs-mh2}
\addtocounter{figure}{-1}
\begin{center}
\setlength{\unitlength}{1cm}
\begin{picture}(15.0,6.5)
\put(-1.5,-0.9)
{\mbox{\epsfysize=7.5cm
 \epsffile{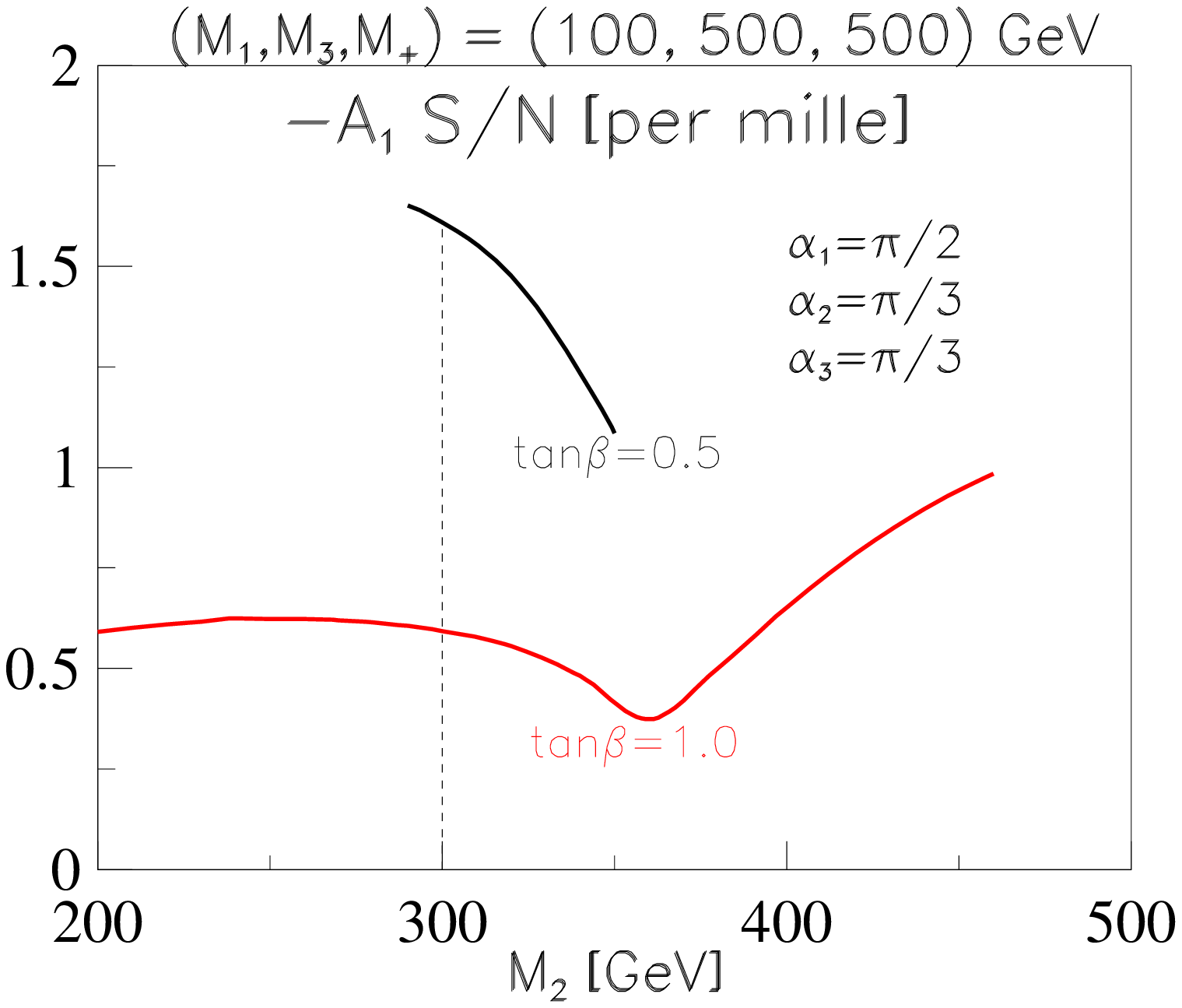}}
 \mbox{\epsfysize=7.5cm
 \epsffile{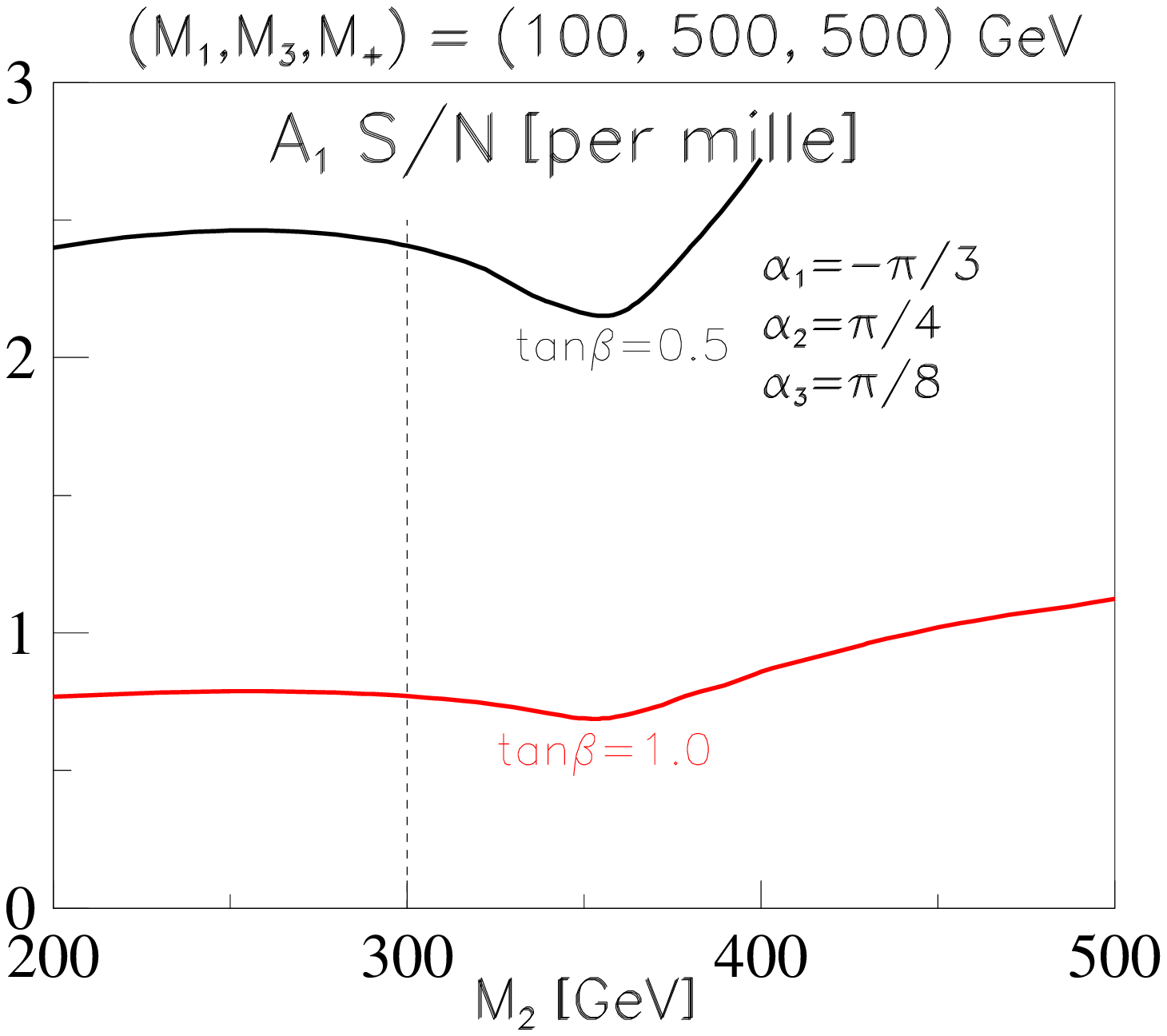}}}
\end{picture}
\vspace*{2mm}
\caption{Signal to noise ratio vs.\  $M_2$, otherwise parameter Set~A.
Rotation matrices $R$ as in Fig.~\ref{Fig:a1-s}.
Dashed line: value of $M_2$ in parameter Set~A.}
\end{center}
\vspace*{-4mm}
\end{figure}

In Fig.~\ref{Fig:a1-s-vs-mh2} we show, for the same rotation matrices as in
Fig.~\ref{Fig:a1-s}, how $S/N$ varies with $M_2$. Here, some of the curves are
cut off at low or high values of $M_2$, since the model becomes inconsistent
or experimentally excluded, as discussed in
Sect.~\ref{sect:mass-parameters-setA}.  In some cases, an enhanced negative
interference occurs as $M_2\simeq2m_t$.

In this study, the parameter $\mu^2$ was allowed to float, as compared with
parameter Set~A, in order to extend the range of allowed values of $M_2$.  In
all cases, except the right panel, with $\tan\beta=1$, the value of $\mu^2$
had to be adjusted as follows: At low values of $M_2$, a lower value of
$\mu^2$ was needed, and at high values of $M_2$, a higher value was needed.

Finally, in Fig.~\ref{Fig:a1-s-vs-mh3}, we show the variation of $S/N$ with
$M_3$, keeping $M_1$ and $M_2$ fixed. There is a tendency for the value to
increase with $M_3$ (because of reduced destructive interference), but
variations of $M_3$ are only allowed within a rather restricted range (see
again Sect.~\ref{sect:mass-parameters-setA} and
Fig.~\ref{Fig:alphas-three-mh3-mhch}).
\begin{figure}[htb]
\refstepcounter{figure}
\label{Fig:a1-s-vs-mh3}
\addtocounter{figure}{-1}
\begin{center}
\setlength{\unitlength}{1cm}
\begin{picture}(15.0,6.5)
\put(-1.5,-0.9)
{\mbox{\epsfysize=7.5cm
 \epsffile{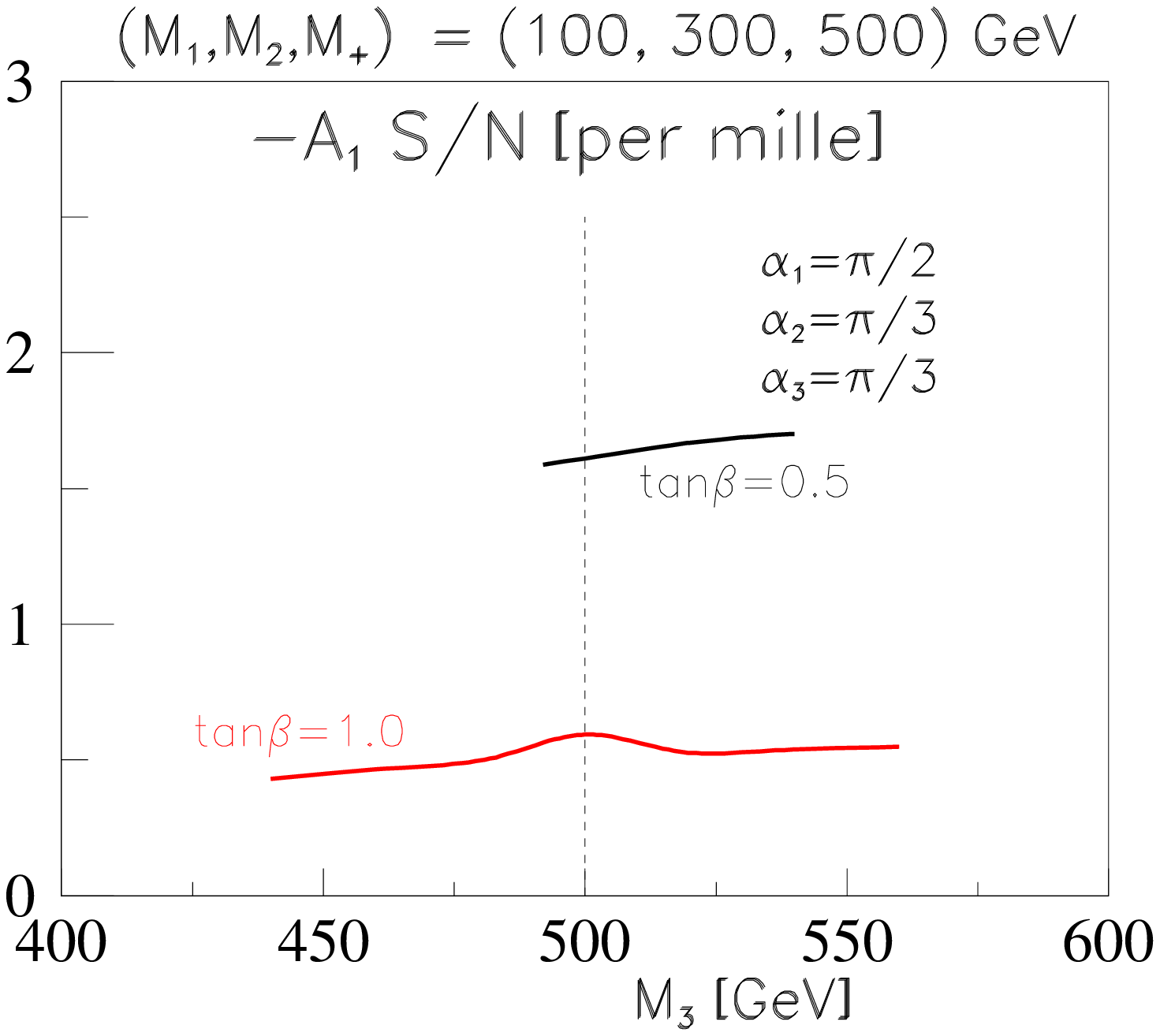}}
 \mbox{\epsfysize=7.5cm
 \epsffile{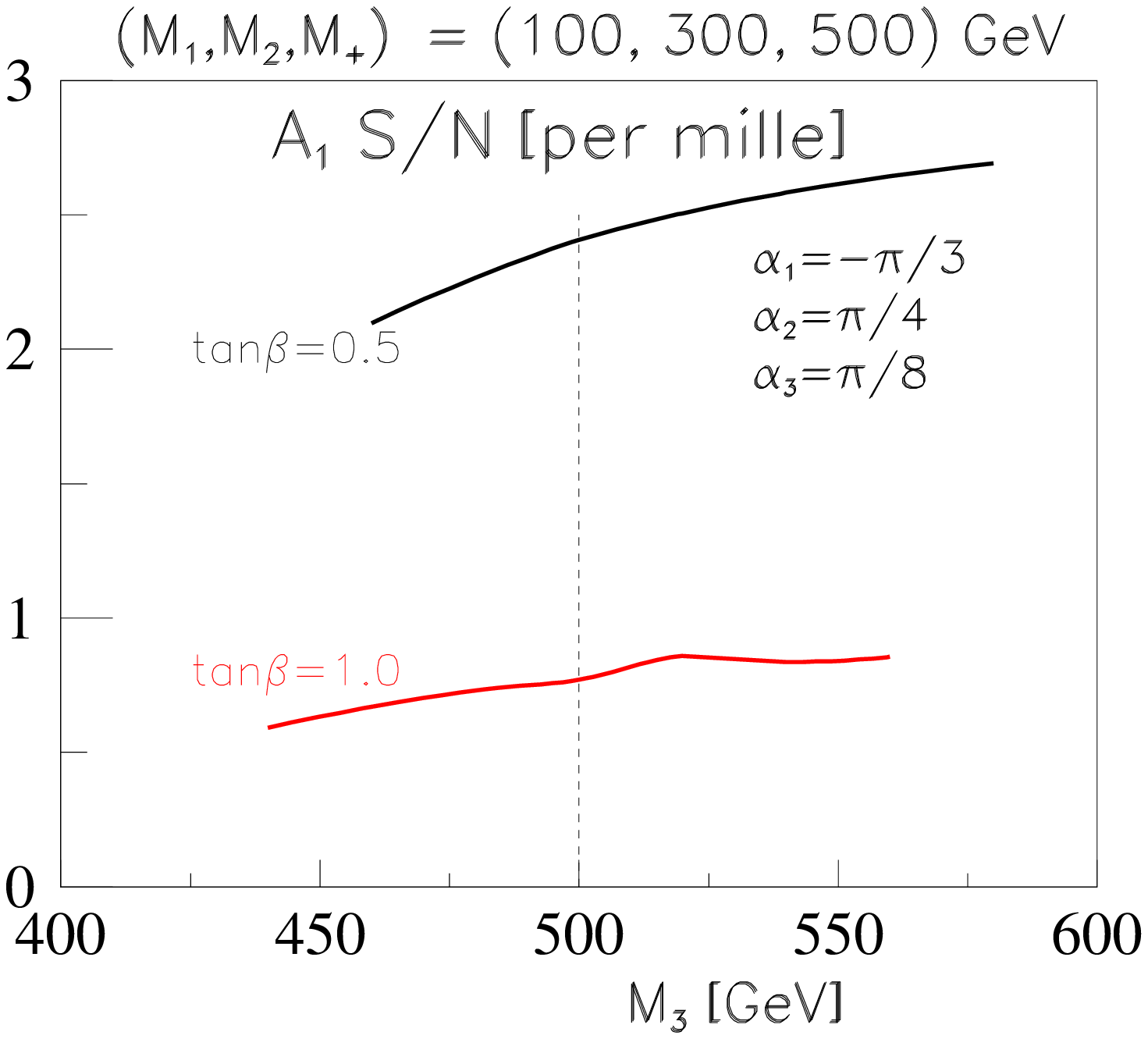}}}
\end{picture}
\vspace*{2mm}
\caption{Signal to noise ratio vs.\  $M_3$, otherwise parameter Set~A.
Rotation angles $\{\alpha_1,\alpha_2,\alpha_3\}$ as indicated.
Dashed line: value of $M_3$ in parameter Set~A.}
\end{center}
\vspace*{-4mm}
\end{figure}

\section{Summary and conclusions}
\label{sect:summary}
\setcounter{equation}{0}
The constraints of stability and tree-level unitarity exclude most of the
multidimensional 2HDM (II) parameter space. Furthermore, the direct searches
(LEP) \cite{Boonekamp:2004ae} and the $\Delta\rho$ \cite{Eidelman:2004wy}
constraints exclude certains domains of the parameters.  Finally, the direct
searches, as well as the $\Delta M_{B_d}$, $R_b$  \cite{Eidelman:2004wy}
and $b\to s\gamma$ \cite{Gambino:2001ew} constraints exclude a light charged
Higgs boson, in particular at low values of $\tan\beta$.
The remaining pockets of allowed parameters will for low values
of $\tan\beta$ allow CP violation in the top-quark
sector, due to the exchange of Higgs bosons that are mixed
with respect to CP.

In order to maximize the CP violation in the top-quark sector that might be
measurable at the LHC, we focus on parameters where
(i) the lightest Higgs boson is rather light, in order
to maximize the relevant loop integrals, and 
(ii) where the product of the CP-violating Yukawa couplings,
parametrized by $\gamma_{CP}^{(1)}$ [see Eq.~(\ref{Eq:gamma_CP-i})]  are large.
The latter constraint requires $\tan\beta$ to be small,
and $|\sin\alpha_1\sin2\alpha_2|$ to be large.

In summary, we note that even in the face of a variety of experimental
constraints, the model is consistent in a number of regions in parameter
space. Apart from exceptional points, these allowed regions yield CP violation
in $t\bar t$ final states produced at the LHC, at a level which can be
explored with a data sample of the order of $10^6$ semileptonic events.

\goodbreak
\bigskip

{\bf Acknowledgments.}  It is a pleasure to thank Okan Camursoy 
and Levent Selbuz for their contributions in the early stages of this work.
This research has been supported in part by the Mission Department of Egypt
and the Research Council of Norway.

\section*{Appendix A} 
\setcounter{equation}{0}
\renewcommand{\thesection}{A}
\setcounter{subsection}{0}
\label{sect:app-stability}
In this Appendix we describe the method used for formulating the necessary and
sufficient conditions for the potential to satisfy stability. 
For earlier approaches to stability, all restricted to simpler potentials,
see \cite{Deshpande:1977rw,Nie:1998yn,Kanemura:1999xf,Ferreira:2004yd}.
We start by rewriting the Higgs doublets as:
\begin{align}
\label{Eq:normphi_i}
\Phi_1=||\Phi_1||{\hat{\Phi}}_1,\hspace*{1cm}
\Phi_2=||\Phi_2||{\hat{\Phi}}_2
\end{align}
where $||\Phi_i||$ is the norm of the spinor $\Phi_i$, 
and ${\hat{\Phi}}_i$ is a unit spinor. By $SU(2)$
invariance, only four combinations of fields may appear:
\begin{equation}
\Phi_1^\dagger\Phi_1=||\Phi_1||^2, \quad 
\Phi_2^\dagger\Phi_2=||\Phi_2||^2, \quad
\Phi_2^\dagger\Phi_1=||\Phi_1||\cdot||\Phi_2||
\left({\hat{\Phi}}_2^\dagger\cdot{\hat{\Phi}}_1\right), \quad
\Phi_1^\dagger\Phi_2=[\Phi_2^\dagger\Phi_1]^*
\nonumber
\end{equation}
We let the norms $||\Phi_i||$ of eq.~(\ref{Eq:normphi_i}) be parametrized as
follows:
\begin{equation}
\label{Eq:parametrization}
||\Phi_1||=r\cos\gamma, \qquad
||\Phi_2||=r\sin\gamma.
\end{equation}
The complex product ${\hat{\Phi}}_2^\dagger\cdot{\hat{\Phi}}_1$
between the two unit spinors will be a complex number $x+iy$ with
$|x+iy|\leq 1$.

Using this parametrization, we can write:
\begin{align}
\Phi_1^\dagger\Phi_1&=r^2\cos^2\gamma, \quad 
\Phi_2^\dagger\Phi_2=r^2\sin^2\gamma, \nonumber \\
\Phi_2^\dagger\Phi_1&=r^2\cos\gamma\sin\gamma
(x+iy), \quad
\Phi_1^\dagger\Phi_2=r^2\cos\gamma\sin\gamma(x-iy),
\end{align}
where $r\geq0$, $\gamma\in[0,\pi/2]$ and $x^2+y^2\leq1$.

The potential can now be written as
\begin{equation}
V=r^4V_4+r^2V_2,
\end{equation}
with the quartic part:
\begin{eqnarray}
\label{Eq:pot_4}
V_4&=&\lambda_1A_1+\lambda_2A_2+\lambda_3A_3+\lambda_4A_4
+\Re\lambda_5A_5+\Im\lambda_5A_6\nonumber\\
&&+\Re\lambda_6A_7+\Im\lambda_6A_8
+\Re\lambda_7A_9+\Im\lambda_7A_{10}
\end{eqnarray}
where
\begin{align}
\label{Eq:A_i}
A_1&=\frac{1}{2}\cos^4\gamma, \quad
A_2=\frac{1}{2}\sin^4\gamma, \quad
A_3=\cos^2\gamma\sin^2\gamma, \nonumber\\
A_4&=(x^2+y^2)\cos^2\gamma\sin^2\gamma,\quad
A_5=(x^2-y^2)\cos^2\gamma\sin^2\gamma,\quad
A_6=2 xy\cos^2\gamma\sin^2\gamma,\nonumber \\
A_7&=2 x\cos^3\gamma\sin\gamma, \quad
A_8=2 y\cos^3\gamma\sin\gamma,\nonumber \\
A_9&=2 x\cos\gamma\sin^3\gamma,\quad
A_{10}=2 y\cos\gamma\sin^3\gamma.
\end{align}
By introducing polar coordinates $x=\rho\cos\theta$ and $y=\rho\sin\theta$,
these coefficients can be written
\begin{align}
\label{Eq:A_i2}
A_1&=\frac{1}{2}\cos^4\gamma, \quad
A_2=\frac{1}{2}\sin^4\gamma, \quad
A_3=\cos^2\gamma\sin^2\gamma, \nonumber\\
A_4&=\rho^2\cos^2\gamma\sin^2\gamma,\quad
A_5=\rho^2\cos^2\gamma\sin^2\gamma\cos(2\theta),\quad
A_6=\rho^2\cos^2\gamma\sin^2\gamma\sin(2\theta),\nonumber \\
A_7&=2\rho\cos^3\gamma\sin\gamma\cos\theta, \quad
A_8=2\rho\cos^3\gamma\sin\gamma\sin\theta,\nonumber \\
A_9&=2\rho\cos\gamma\sin^3\gamma\cos\theta,\quad
A_{10}=2\rho\cos\gamma\sin^3\gamma\sin\theta,
\end{align}
where 
$\gamma\in[0,\pi/2]$, $\rho\in[0,1]$ and $\theta\in[0,2\pi\rangle$.

\subsection{Symmetries of $V_4$}
We note some  symmetries of the {\it quartic} potential under the
parametrization (\ref{Eq:parametrization}). 
They are conditional symmetries, where the potential
is invariant with some ``compensating'' interchange of $\lambda$s:

\noindent{\bf Symmetry I:} \\
The interchange
\begin{equation}
y\leftrightarrow -y\quad \text{or}\quad
\theta\leftrightarrow2\pi-\theta
\end{equation}
can be compensated by
\begin{equation}
\{\Im\lambda_5,\; \Im\lambda_6,\; \Im\lambda_7\} \leftrightarrow
\{-\Im\lambda_5,\; -\Im\lambda_6,\; -\Im\lambda_7\}
\end{equation}
This is of course nothing but the reality condition of the potential.

\noindent{\bf Symmetry II:} \\
Under
\begin{equation}
\gamma\leftrightarrow\frac{\pi}{2}-\gamma
\end{equation}
together with
\begin{equation}
\lambda_1\leftrightarrow \lambda_2, \quad \text{and}\quad
\lambda_6\leftrightarrow \lambda_7
\end{equation}
the potential is invariant.
This is the symmetry under the interchange $\Phi_1\leftrightarrow\Phi_2$.

\noindent{\bf Symmetry III:} \\
Under
\begin{equation}
(x,y)\leftrightarrow(-x,-y)\quad\text{or}\quad 
\theta\leftrightarrow\pi+\theta\mod 2\pi
\end{equation}
together with
\begin{equation}
\{\lambda_6,\lambda_7\}\leftrightarrow \{-\lambda_6,-\lambda_7\}
\end{equation}
the potential is invariant.  This is related to the well-known $Z_2$ symmetry
\cite{Weinberg:1976hu,Glashow:1976nt}.

\noindent{\bf Symmetry IV:} \\
Under
\begin{equation}
(x,y)\leftrightarrow(y,x)\quad\text{or}\quad 
\theta\leftrightarrow\frac{\pi}{2}-\theta\mod 2\pi
\end{equation}
together with
\begin{equation}
\Re\lambda_5\leftrightarrow-\Re\lambda_5\quad\text{and}\quad
\{\Re\lambda_6,\Re\lambda_7\}\leftrightarrow \{\Im\lambda_6,\Im\lambda_7\}
\end{equation}
the potential is invariant.

\subsection{Stability}
For the stability condition to be satisfied, $V_4$ must be positive for
all combinations of $\gamma\in[0,\pi/2]$, $\rho\in[0,1]$ and 
$\theta\in[0,2\pi\rangle$. This is both a necessary and a
sufficient condition. Whenever $\lambda_4\leq0$, the potential will have
its global maximum and minimum when $\rho=1$. Thus, it is sufficient to 
check that $V_4(\gamma,\theta;\rho=1)$ satisfies stability when
$\lambda_4$ is non-positive. In order to see this, we return to eqs.
(\ref{Eq:pot_4}) and (\ref{Eq:A_i}) and rewrite the potential as
\begin{equation}
V_4=\lambda_4(x^2+y^2)\cos^2\gamma\sin^2\gamma+h(x,y)\nonumber
\end{equation}
where $h(x,y)$ is a harmonic function. The maximum principle tells us
that $h(x,y)$ will attain its global minimum at the boundary where
$x^2+y^2=1$ $(\rho=1)$. Whenever $\lambda_4\leq0$, the term 
$\lambda_4(x^2+y^2)\cos^2\gamma\sin^2\gamma$ will also attain its
minimum whenever $\rho=1$, and so will $V_4$.

Some points from the parameter space give us some rather simple
stability conditions. We now turn our attention towards these special 
points.

\noindent{\underline{\bf $\gamma=0$ or $\gamma=\pi/2$}}\\
First we consider the boundary points $\gamma=0$ and $\gamma=\pi/2$.
\begin{eqnarray}
V_4(\gamma=0)&=&\frac{\lambda_1}{2}\nonumber\\
V_4(\gamma=\pi/2)&=&\frac{\lambda_2}{2}\nonumber
\end{eqnarray}
This leaves us with the ``trivial'' stability conditions of
(\ref{Eq:pos-simple}):
\begin{equation}\label{Eq:trivialstability}
\lambda_1>0\quad {\rm and}\quad \lambda_2>0.
\end{equation}

\noindent{\underline{\bf $\rho=0$}}\\
Considering the points where $\rho=0$,
we find that
\begin{equation}
\label{Eq:gammavalues}
V_4(\rho=0)=\frac{\lambda_1}{2}\cos^4\gamma
+\frac{\lambda_2}{2}\sin^4\gamma+\lambda_3\cos^2\gamma\sin^2\gamma.
\end{equation}
Thus, we must require that
\begin{equation}
\lambda_3>-\frac{1}{2}
\left(\frac{\lambda_1}{\tan^2\gamma}+\lambda_2\tan^2\gamma\right).
\end{equation}
The right-hand side has its minimum for
$\tan^2\gamma=\sqrt{\lambda_1/\lambda_2}$, and we obtain the
following necessary constraint on $\lambda_3$:
\begin{equation}
\label{Eq:min-lambda3}
\lambda_3>-\sqrt{\lambda_1\lambda_2},
\end{equation}
in agreement with \cite{Deshpande:1977rw}.
\subsection{The limit $\lambda_6=\lambda_7=0$}
It is instructive to consider the simple limit
\begin{equation}
\lambda_6=\lambda_7=0
\end{equation}
Then the quartic part of the potential can be written
\begin{align}
V_4=&\frac{\lambda_1}{2}\cos^4\gamma
+\frac{\lambda_2}{2}\sin^4\gamma
+[\lambda_3
+\rho^2(\lambda_4+\Re\lambda_5\cos2\theta+\Im\lambda_5\sin2\theta)]
\cos^2\gamma\sin^2\gamma
\end{align}
This expression has the same structure as (\ref{Eq:gammavalues}),
with 
\begin{equation}
\label{Eq:lambda_3tilde}
\lambda_3\to
\lambda_3+\rho^2[\lambda_4+\Re\lambda_5\cos2\theta+\Im\lambda_5\sin2\theta].
\end{equation}
Thus, the condition for stability can be adapted from (\ref{Eq:min-lambda3}),
leading to:
\begin{equation} \label{Eq:lambda45-pos}
\lambda_3+\min[0,\lambda_4-|\lambda_5|]
>-\sqrt{\lambda_1\lambda_2},
\end{equation}
as obtained by \cite{Deshpande:1977rw}.
However, we stress that this constraint only applies when
$\lambda_6=\lambda_7=0$.
\section*{Appendix B} \label{sect:app-couplings}
\setcounter{equation}{0}
\renewcommand{\thesection}{B}
\setcounter{subsection}{0}
The Higgs--vector-boson couplings can be extracted from the covariant
derivatives:
\begin{align}
&(D^\mu \Phi_1)^\dagger (D_\mu \Phi_1) 
+ (D^\mu \Phi_2)^\dagger (D_\mu \Phi_2) \nonumber \\
&=\frac{ig}{2\cos\theta_W}Z^\mu 
\{-\cos2\theta_W(\varphi_1^- \partial_\mu \varphi_1^+ 
- \varphi_1^+ \partial_\mu \varphi_1^-
+ \varphi_2^- \partial_\mu \varphi_2^+
- \varphi_2^+ \partial_\mu \varphi_2^-) \nonumber \\
&+i[\eta_1 \partial_\mu \chi_1 - \chi_1 \partial_\mu \eta_1
   +\eta_2 \partial_\mu \chi_2 - \eta_2 \partial_\mu \chi_2]\} \nonumber \\
&-\half ig W^\mu{}^\dagger 
[(\eta_1-i\chi_1)\partial_\mu \varphi_1^+ 
- \partial_\mu (\eta_1-i\chi_1) \varphi_1^+
+ (\eta_2-i\chi_2)\partial_\mu \varphi_2^+
- \partial_\mu (\eta_2-i\chi_2) \varphi_2^+]
 \nonumber \\
&+\half ig W^\mu
[ (\eta_1+i\chi_1) \partial_\mu \varphi_1^-
- \varphi_1^- \partial_\mu (\eta_1+i\chi_1) 
+ (\eta_2+i\chi_2) \partial_\mu \varphi_2^-
- \varphi_2^- \partial_\mu (\eta_2+i\chi_2) ] +\cdots
\end{align}

These Higgs fields have to be transformed into the physical basis:
\begin{equation}
\eta_i=R_{ji}H_j, \quad i=1,2,3,
\end{equation}
with
\begin{equation}
\begin{pmatrix}
\varphi_1^\pm \\
\varphi_2^\pm 
\end{pmatrix}
=\begin{pmatrix}
\cos\beta & -\sin\beta \\
\sin\beta & \cos\beta
\end{pmatrix}
\begin{pmatrix}
G^\pm \\
H^\pm 
\end{pmatrix}, \quad
\begin{pmatrix}
\chi_1 \\
\chi_2 
\end{pmatrix}
=\begin{pmatrix}
\cos\beta & -\sin\beta \\
\sin\beta & \cos\beta
\end{pmatrix}
\begin{pmatrix}
G^0 \\
\eta_3
\end{pmatrix}
\end{equation}

With all momenta incoming (in an obvious notation), we find
\begin{alignat}{2}
&ZH^+H^-: &\qquad
&\frac{-ig\cos2\theta_W}{2\cos\theta_W}(p_\mu^+-p_\mu^-), \nonumber \\
&ZG^+G^-: &\qquad
&\frac{-ig\cos2\theta_W}{2\cos\theta_W}(p_\mu^+-p_\mu^-), \nonumber \\
&ZH_j H_k: &\qquad
&\frac{-g}{2\cos\theta_W}
[(\sin\beta R_{j1}-\cos\beta R_{j2})R_{k3}
-(\sin\beta R_{k1}-\cos\beta R_{k2})R_{j3}]
(p_\mu^j-p_\mu^k),  \nonumber \\
&ZH_j G^0: &\qquad
&\frac{g}{2\cos\theta_W}
(\cos\beta R_{j1}+\sin\beta R_{j2})(p_\mu^j-p_\mu^0),
\end{alignat}
and
\begin{alignat}{2}
&W^\pm H^\mp H_j: &\qquad
&\frac{g}{2}
[\pm i(\sin\beta R_{j1}-\cos\beta R_{j2})+ R_{j3}]
(p_\mu^j-p_\mu^\mp), \nonumber \\
&W^\pm G^\mp H_j: &\qquad
&\frac{\mp ig}{2}(\cos\beta R_{j1}+\sin\beta R_{j2})
(p_\mu^j-p_\mu^\mp), \nonumber \\
&W^\pm G^\mp G^0: &\qquad
&\frac{g}{2}(p_\mu^0-p_\mu^\mp)
\end{alignat}
There are no $Z H^\pm G^\mp$ or $W^\pm H^\mp G^0$ couplings.
The CP-conserving limit is obtained by evaluating $R$
for $\alpha_2=0$, $\alpha_3=0$, $\alpha_1=\alpha+\pi/2$, with the mapping
$H_1\to h$, $H_2\to -H$ and $H_3\to A$.
In that limit, we recover the results of \cite{HHG}.

\end{document}